%Paper: cond-mat/9511126
%From: Komineas Stavros <komineas@iesl.forth.gr>
%Date: Mon, 27 Nov 95 17:09:35 +0200
%Date (revised): Wed, 29 Nov 95 11:06:02 +0200

\language1
\mag=\magstep1
\count0=1
\font\bsym=cmbsy10
\def\bnabla{\hbox{$\textfont2=\bsym \nabla$}}
\font\se=cmr7
\font\huge=cmr12
\font\biggb=cmbx10 scaled \magstep3
\font\ini=cmbx12
\def\nb #1{{\hbox{\bf #1}}}
\hsize=15truecm
\vsize=23truecm
\hoffset=0.6 truecm
\tolerance=10000
\hbadness=10000
\hfuzz=20pt

\vbox{\vskip 1.5truecm}
\centerline{\biggb Topology and dynamics in ferromagnetic media}

\vbox{\vskip 2truecm}

\centerline{\huge S. Komineas and N. Papanicolaou}
\centerline{Department of Physics, University of Crete}
\centerline{and Research Center of Crete}
\centerline{Heraklion, Greece}

\vbox{\vskip 5truecm}

\vfill

\centerline{\bf Abstract}
\bigskip
{A direct link between the topological complexity of ferromagnetic
media and their dynamics has recently been established through the
construction of unambiguous conservation laws as moments of a topological
vorticity. In the present paper we carry out this program under completely
realistic conditions, with due account of the long-range magnetostatic
field and related boundary effects. In particular, we derive unambiguous
expressions for the linear and angular momentum in a ferromagnetic film
which are then used to study the dynamics of magnetic bubbles under the
influence of an applied magnetic-field gradient. The semi-empirical golden
rule of bubble dynamics is verified in its gross features but not in its finer
details. A byproduct of our analysis is a set of virial theorems generalizing
Derrick's scaling relation as well as a detailed recalculation of the
fundamental magnetic bubble.}

\vfill
\eject

\noindent
{\ini I.\ Introduction}
\bigskip
Magnetic bubbles have been known to exhibit some distinct dynamical features
due to their nontrivial topological structure. The inherent link between
topology and dynamics was already apparent in the early work of Thiele [1]
as well as in many investigations that followed [2, 3]. The essence of the
early work is best summarized by the experimentally observed skew
deflection of magnetic bubbles under the influence of an applied magnetic-field
gradient. The so-called golden rule of bubble dynamics relates the
deflection angle $\delta$ to the winding number $Q$ by
$${gr^2\over 2V}\sin\delta =Q,\eqno (1.1)$$
where $g$ is the strength of the applied field gradient, $r$ is the
bubble radius and $V$ its speed. Relation (1.1) is remarkable in two respects.
First, it suggests that only topologically trivial $(Q=0)$ bubbles
move in the direction of the gradient $(\delta =0)$, even though such a
behavior would naively be expected for all magnetic bubbles; in fact, bubbles
with a nonvanishing winding number $(Q=\pm 1,\, \pm 2,\, \ldots)$ tend to
be deflected in a direction nearly perpendicular $(\delta\sim 90^{\circ})$
to the applied gradient. Second, Eq. (1.1) implies some sort of a topological
quantization in that it relates the integer-valued winding number to
experimentally measured quantities that can, in principle, assume any values.

Although golden rule (1.1) has been employed with considerable success in
the analysis of actual experiments, especially for hard
$(|Q|\gg 1)$ bubbles, it is still a semi-empirical relation whose precise
meaning and domain of validity must be specified. For instance, the meaning
of the various quantities entering Eq. (1.1) needs to be explained because
magnetic bubbles are extended structures rather than point-like particles.
Furthermore the usual derivations of (1.1) are based on the assumption that
the bubble reaches a steady state, in the presence of dissipation, in which
the deflection angle $\delta$, the radius $r$ and the speed $V$ approach
constant values. But such an assumption was never justified and is
actually incorrect. In practice, experiments are analyzed by applying Eq. (1.1)
with average values for the deflection angle and the speed and by assuming
that the radius does not change significantly during the application of the
gradient.

In some recent work on this subject [4] the link between topology and dynamics
was made explicit through the construction of unambiguous conservation
laws as moments of a suitable topological vorticity. The important
qualitative features of bubble dynamics became then apparent. Thus, in the
absence of external magnetic-field gradients or other perturbations,
bubbles with a nonvanishing winding number cannot move freely but are
always spontaneously pinned. On the other hand, in the absence of dissipation,
a bubble would be deflected at a right angle $(\delta =90^{\circ})$ with
respect to an applied magnetic-field gradient, with a drift velocity that
can be calculated analytically in some important special cases [4] and is
generally consistent with Eq. (1.1). The emerging picture is thus
analogous to the Hall motion of an electron as well as to the Magnus
effect of fluid dynamics. These analogies further suggest that the
deflection angle should deviate from $90^{\circ}$ in the presence of
dissipation. However an exact calculation of the deflection angle, i.e.,
a derivation of the golden rule, is no longer possible on the basis of
conservation laws alone.

Therefore the semi-quantitative picture derived from the conservation laws
must be supplemented by some results from a numerical solution of the
underlying Landau-Lifshitz equation. Such a solution is not
straightforward under completely realistic conditions; calculation of
the long-range magnetostatic field is always a problem and the finite
thickness of actual magnetic films forces one to work with a three-dimensional
(3D) grid, even though the essential topological structure of magnetic
bubbles is two-dimensional (2D). Hence our numerical efforts have thus
far been restricted to strictly 2D models of increasing complexity [5, 6].

It is the purpose of the present paper to provide a precise formulation
within the quasi-2D geometry of realistic ferromagnetic films, taking into
account the effects from the film boundaries and the magnetostatic field.
In Section II we review some basic facts about the Landau-Lifshitz equation and
introduce convenient (rationalized) physical units. In Section III we discuss
the two ingredients that are important to establish a direct link between
the topological complexity of magnetic structures and their dynamics, the
gyrovector and the stress tensor. The derivation of unambiguous conservation
laws is then carried out in Section IV in the presence of film boundaries.
A byproduct of our study of conservation laws is a set of virial
theorems that generalize the well-known scaling relation of Derrick [7],
an issue taken up in Section V. A detailed numerical calculation of the
fundamental $(Q=1)$ bubble is presented in Section VI which is consistent
with the virial theorems. The issue of skew deflection in an applied
magnetic-field gradient is studied in Section VII where we find that the
semi-empirical golden rule is verified in its gross features but not in its
details. Our conclusions are summarized in Section VIII together with some
suggestions for further work.

\bigskip
\bigskip
\bigskip
\noindent
{\ini II.\ The Landau-Lifshitz equation}
\bigskip
A ferromagnetic medium is described in terms of the density of magnetic
moment or magnetization ${\nb M}$  which is due primarily to the electron
spins but may include contributions also from the orbital motion. In general,
the vector ${\nb M}=(M_1, M_2, M_3)$ is a function of position and time except
that its magnitude is nearly constant for a wide temperature range
sufficiently below the Curie point. Thus we write
$${\nb M}={\nb M}({\nb x}, t),\qquad {\nb M}^2\equiv M_1^2+M_2^2+M_3^2=
M_0^2,\eqno (2.1)$$
where ${\nb x}=(x_1, x_2, x_3)$ is the position vector, $t$ is the time
variable, and the constant $M_0$ is the saturation magnetization.

Static as well as dynamical properties of the magnetization are governed
by the Landau-Lifshitz equation
$${\partial {\nb M}\over\partial t}+\gamma ({\nb M}\times {\nb F})=
{\lambda\over M_0}
\left( {\nb M}\times{\partial {\nb M}\over\partial t}\right) ,\eqno (2.2)$$
which describes precession around an effective field ${\nb F}$ with the
constant $\gamma$ given by
$$\gamma ={g_e |e|\over 2m_ec},\eqno (2.3)$$

\noindent
where $g_e\sim 2$ is the gyromagnetic ratio, $e$ the electron charge,
$m_e$ the electron mass, and $c$ the velocity of light. Equation (2.2) also
includes a phenomenological (Landau-Gilbert) dissipative term where the
dissipation constant $\lambda$ is dimensionless. This choice of dissipation
preserves the magnitude of magnetization.

The effective field ${\nb F}$ may be written as
$${\nb F}={\nb F}_e+{\nb F}_a+{\nb H}_b+{\nb H}.\eqno (2.4)$$
Here ${\nb F}_e$ is the exchange field
$${\nb F}_e={2A\over M_0^2}\Delta {\nb M},\eqno (2.5)$$
where $A$ is the exchange stiffness constant and $\Delta$ the Laplace operator.
${\nb F}_a$ is the anisotropy field
$${\nb F}_a=-{2K\over M_0^2}(M_1, M_2, 0),\eqno (2.6)$$
where $K$ is a positive constant leading to an easy axis in the third
direction. In ferromagnetic films made out of bubble materials the easy
axis is perpendicular to the film surface [2]. ${\nb H}_b$ is a
uniform bias field,
$${\nb H}_b=(0, 0, H_b),\qquad H_b=\hbox{const},\eqno (2.7)$$
applied along the easy axis. Finally ${\nb H}$ is the magnetic field
produced by the magnetization itself and thus satisfies the magnetostatic
equations
$$\bnabla\times{\nb H}=0,\qquad \bnabla\cdot{\nb B}=0;\qquad
{\nb B}=\nb H+4\pi\nb M,\eqno (2.8)$$
where $\nb B$ is the corresponding magnetic induction. The use of the
magnetostatic instead of the complete Maxwell equations is justified by the
fact that time variations of magnetic structures of practical interest
are slow.

Numerical values of the various constants introduced above may be found
in Ref. [2] for a number of ferromagnetic materials. However using all these
constants within a theoretical development clouds the underlying simplicity
of the Landau-Lifshitz equation. Hence we introduce rationalized physical
units as follows. First, we work with the normalized magnetization
$$\nb m=\nb M/M_0,\qquad \nb m^2=1.\eqno (2.9)$$
Second, we measure distance, time and magnetic field (induction) in units of
$$\sqrt{{A\over 2\pi M^2_0}},\qquad {1\over 4\pi\gamma M_0}\quad
\hbox{and}\quad 4\pi M_0,\eqno (2.10)$$

\noindent
respectively. We further define the dimensionless anisotropy constant
$$\kappa ={K\over 2\pi M^2_0},\eqno (2.11)$$
which is usually referred to as the quality factor. Finally, we introduce
new symbols for dimensionless magnetic fields, such as
$\nb h=\nb H/4\pi M_0$, but maintain the same symbols $\nb x$ and $t$ for
the rationalized space and time variables.

The Landau-Lifshitz equation is then written as
$$\dot{\nb m}+(\nb m\times\nb f)=\lambda (\nb m\times\dot{\nb m}),\qquad
\nb m^2=1,\eqno (2.12)$$
where the dot denotes differentiation with respect to time, a convention
that will be adopted from now on. The effective field $\nb f$ is given by
$$\nb f=\Delta\nb m-\kappa (m_1, m_2, 0)+\nb h_b+\nb h,\eqno (2.13)$$
where $\nb h_b=(0, 0, h_b)$ is the bias field and $\nb h$ satisfies the
magnetostatic equations
$$\bnabla\times\nb h=0,\qquad \bnabla\cdot\nb b=0;\qquad
\nb b=\nb h+\nb m.\eqno (2.14)$$
The only free parameters are now the quality factor $\kappa$, the dissipation
constant $\lambda$, and the bias field $h_b$. One should add that Eq. (2.12)
differs from the Landau-Lifshitz equation studied in our earlier work [4] by
the overall replacement $\nb m\to -\nb m$ that originates in a sign difference
between magnetization and spin introduced by the negative electron charge.
The resulting sign differences at various stages of the calculation will
be incorporated without further notice.

The remainder of this section is devoted to a discussion of the hamiltonian
structure associated with the Landau-Lifshitz equation at vanishing
dissipation $(\lambda =0)$. Then we write
$$\dot{\nb m}+(\nb m\times\nb f)=0,\qquad \nb m^2=1,\eqno (2.15)$$
where the effective field $\nb f$ is still given by Eq. (2.13) and may be
expressed entirely in terms of the magnetization once the magnetostatic
field $\nb h$ is determined by solving the linear system (2.14). Now a
conserved energy functional $W=W(\nb m)$ exists such that the effective
field is obtained through the general relation
$$\nb f =-{\delta W\over \delta \nb m},\eqno (2.16)$$
where the symbol $\delta$ denotes the usual functional derivative.
Equation (2.16) together with Eq. (2.15) imply that the functional $W$ is
indeed conserved, and that (2.15) is the Hamilton equation associated with
the hamiltonian $W$ endowed with the Poisson-bracket relations
$$\{ m_i(\nb x), m_j(\nb x')\} =\varepsilon_{ijk}m_k(\nb x)
\delta (\nb x-\nb x'),\eqno (2.17)$$
which are reminiscent of the spin commutation relations. Here
$\varepsilon_{ijk}$ is the 3D antisymmetric tensor and the usual summation
convention for repeated indices is invoked.

In order to display the explicit form of the energy functional we write
$$W=W_e+W_a+W_b+W_m,\eqno (2.18)$$
where the four terms correspond to the exchange, anisotropy, bias, and
mangetostatic field. The more or less obvious choice of the exchange energy,
$$W_e=\int w_edV,\qquad w_e={1\over 2}(\partial_i\nb m\cdot\partial_i\nb m),
\eqno (2.19)$$
where $w_e$ is the corresponding energy density, requires some qualification
in the presence of boundaries. Thus we consider the functional variation
$$\delta W_e=\int(\partial_i\delta\nb m\cdot\partial_i\nb m)dV=\oint
(\delta\nb m\cdot\partial_i\nb m)dS_i-\int
(\delta\nb m\cdot\Delta\nb m)dV,\eqno (2.20)$$
where the surface-element vector $d\nb S=(dS_1, dS_2, dS_3)$ is
perpendicular to the boundaries of the ferromagnetic medium. Equation (2.20)
would yield the desired relation
$$\nb F_e=-{\delta W_e\over \delta\nb m}=\Delta \nb m,\eqno (2.21)$$
if the surface integral were absent; that is, if the gradient of the
magnetization along the normal to the surface vanished. We write
symbolically
$${\partial \nb m\over\partial n}=0,\eqno (2.22)$$
which will be viewed as a boundary condition to be imposed at the free
boundaries of the medium, in addition to the familiar boundary conditions
of magnetostatics. This ``unpinned'' boundary condition was previously
employed in the study of ferromagnetic films [2] and will play an
important role in the following.

On the other hand, the usual bulk expressions for the anisotropy and bias
(Zeeman) energies,
$$\eqalign{W_a=\int w_adV,&\qquad w_a={\kappa\over 2}(m_1^2+m_2^2)\cr
\noalign{\medskip}
W_b=\int w_bdV,&\qquad w_b=h_b(1-m_3),\cr}\eqno (2.23)$$
are free of boundary ambiguities and obviously yield the corresponding
contributions to the effective field through the general relation (2.16).
Note that in the Zeeman energy we have subtracted the (trivial) contribution
from the state
$$\nb m_0=(0, 0, 1),\eqno (2.24)$$
which describes a fully saturated ferromagnet and will thus be referred to
as the ground state. Configuration (2.24) is the simplest example of a static
solution of the Landau-Lifshitz equation.

Boundary effects are also important in the definition of the magnetostatic
energy and will be described here in some detail. We begin with the reasonable
ansatz
$$W_m={1\over 2}\int\nb h^2dV,\eqno (2.25)$$
where it is understood that the integral extends over all volume, inside and
outside the material, and that the field $\nb h$ is expressed in terms of
the magnetization through Eqs. (2.14). However, in order to justify that
(2.25) is an appropriate choice of the mangetostatic energy within the context
of the Landau-Lifshitz equation, one must show that
$$\nb h=-{\delta W_m\over\delta\nb m}.\eqno (2.26)$$
Such a demonstration is not completely straightforward because of the
implicit dependence of $\nb h$ on the magnetization.

To make this dependence explicit we introduce a scalar potential $\psi$ from
$$\nb h=-\bnabla\psi ,\qquad \Delta\psi =(\bnabla\cdot\nb m),\eqno (2.27)$$
and solve the Poisson equation to obtain
$$\psi (\nb x)={1\over 4\pi}\left[\oint {\nb m(\nb x')\cdot d\nb S'\over
|\nb x-\nb x'|}-\int{(\bnabla\cdot\nb m)(\nb x')\over |\nb x-\nb x'|}
dV'\right] ,\eqno (2.28)$$
where the surface integral extends over the boundaries of the ferromagnetic
medium, if any, and the volume integral over the bulk of the medium. Applying
a careful partial integration yields the equivalent relation
$$\psi (\nb x)={1\over 4\pi}\int{(\nb x-\nb x')\cdot\nb m (\nb x')
\over |\nb x-\nb x'|^3}dV',\eqno (2.29)$$
whose advantage is that it contains no derivatives of the magnetization and is
valid irrespectively of the presence of boundaries.

As an elementary illustration we consider a ferromagnetic film of thickness
$d$ (see Fig. 1) and assume that the magnetization is equal to its
ground-state value (2.24) inside the film (region I) and vanishes outside
(regions II and III). An explicit calculation of the integral in
Eq. (2.29) then yields
$$\psi =\cases{d/2, &$\qquad x_3>d/2$\cr
\noalign{\smallskip}
x_3, &$-d/2<x_3<d/2$\cr
\noalign{\smallskip}
-d/2, &$\qquad x_3<-d/2.$\cr}\eqno (2.30)$$
Therefore the magnetic field is given by $\nb h=-\bnabla\psi =-\nb m_0$
inside the film and vanishes outside. The magnetic induction
$\nb b=\nb h+\nb m_0$ vanishes everywhere.

We now return to the magnetostatic energy (2.25) and replace $\nb h^2$
by $-\nb h\cdot (\bnabla\psi)$. An application of the divergence
theorem and Eqs. (2.14) then gives
$$W_m={1\over 2}\left[\oint\psi (\nb m\cdot d\nb S)-
\int\psi (\bnabla\cdot\nb m)dV\right] ,\eqno (2.31)$$
where we have also used the fact that $\psi$ is continuous across the
boundary and that the difference between the normal components of the
magnetic field on the two sides of the boundary is equal to the normal
component of the magnetization. A further partial integration
transforms (2.31) into
$$W_m=-{1\over 2}\int (\nb h\cdot\nb m)dV,\eqno (2.32)$$
which shares with Eq. (2.29) the property that it is valid whether
or not boundaries are present. Hence, using this form of the
magnetostatic energy and a magnetic field calculated from Eq. (2.29), the
basic relation (2.26) is established by straightforward manipulations.

We complete the discussion of the canonical structure noting that the
Landau-Lifshitz equation is actually a constrained hamiltonian system.
Nevertheless one may resolve the constraint $\nb m^2=1$ explicitly
using, for example, the spherical parametrization
$$m_1=\sin\Theta\cos\Phi ,\qquad m_2=\sin\Theta\sin\Phi ,
\qquad m_3=\cos\Theta .\eqno (2.33)$$
The energy functional is then parametrized in terms of the two independent
fields $\Theta$ and $\Phi$ and the general form of the Landau-Lifshitz
equation reads
$$\sin\Theta\,\,\dot{\Theta}=-{\delta W\over\delta\Phi},\qquad
\sin\Theta\,\,\dot{\Phi}={\delta W\over\delta\Theta},\eqno (2.34)$$
which suggests that the pair of fields
$$\Pi =\cos\Theta\quad\hbox{and}\quad\Phi\eqno (2.35)$$
is a canonical pair:
$$\dot{\Pi}={\delta W\over\delta\Phi},\qquad
\dot{\Phi}=-{\delta W\over\delta\Pi}.\eqno (2.36)$$
However most of the special dynamical features of the ferromagnetic
continuum emerge exactly when the definition of the canonical variables
(2.35) encounters ambiguities due to a possibly nontrivial topological
structure of the magnetization.

Finally we return briefly to the issue of dissipation and rewrite
Eq. (2.12) in the equivalent form
$$\eqalign{\dot{\nb m} &+(\nb m\times\nb G)=0,\qquad\nb m^2=1,\cr
\noalign{\medskip}
\nb G &=\lambda_1\nb f+\lambda_2(\nb m\times\nb f),\qquad
\lambda_1={1\over 1+\lambda^2},\qquad
\lambda_2={\lambda\over 1+\lambda^2}.\cr}\eqno (2.37)$$

\noindent
We then examine the rate at which the energy changes in the presence of
dissipation:
$$\dot{W} =\int
\left({\delta W\over\delta\nb m}\cdot\dot{\nb m}\right) dV
=-\lambda_2\int [\nb f^{\,2}-(\nb m\cdot\nb f)^2]dV.\eqno (2.38)$$
Because $\nb m$ is a unit vector the integrand in the last step of Eq. (2.38)
is positive definite and the energy decreases when the dissipation constant
$\lambda$ is positive.

\bigskip
\bigskip
\bigskip
\noindent
{\ini III.\ Gyrovector and the stress tensor}
\bigskip
The key quantity for the description of both topological and dynamical
properties of the magnetization is the gyrovector or vorticity
$\hbox{\bf ã}=(\gamma_1, \gamma_2, \gamma_3)$ whose cartesian components
are given by
$$\gamma_i=-{1\over 2}\varepsilon_{ijk}(\partial_j\nb m\times\partial_k\nb m)
\cdot \nb m.\eqno (3.1)$$
The former terminology was introduced in the early work [1] but the latter
seems more appropriate in view of the significant formal analogy of the
vector $\hbox{\bf ã}$ with ordinary vorticity in fluid dynamics.
Nevertheless one should stress that $\hbox{\bf ã}$ is not related to actual
rotational motion in the ferromagnetic continuum but rather to the
topological complexity of the magnetization. For the moment, we are concerned
with instantaneous properties of the unit vector field $\nb m=\nb m(\nb x)$
at some instant $t$ that is not displayed explicitly. Questions of
dynamics will be addressed later in this section.

An immediate consequence of the definition (3.1) and the constraint
$\nb m^2=1$ is that the vorticity field is solenoidal,
$$\bnabla\cdot\hbox{\bf ã}=0,\eqno (3.2)$$
and hence the corresponding vortex lines cannot terminate but at the
boundaries of the ferromagnetic medium. The precise nature of vortex lines
is revealed by expressing the vorticity in terms of the canonical variables
(2.35),
$$\hbox{\bf ã}=\bnabla\Pi\times\bnabla\Phi ,\eqno (3.3)$$
a relation that suggests an analogy of $\Pi$ and $\Phi$ with the Clebsch
potentials of fluid dynamics [8]. It also establishes that vortex lines are
defined as the intersections of the two surfaces $\Pi (\nb x)=c_1$ and
$\Phi (\nb x)=c_2$ where $c_1$ and $c_2$ are arbitrary constants. In other
words, vortex lines are the curves along which the magnetization vector
$\nb m$ remains constant.

Such a simple definition of vortex lines allows a transparent topological
classification of the possible distributions of magnetization. We shall
consider the physically interesting class of configurations
$\nb m=\nb m(\nb x)$ that are differentiable functions of position and
approach the ground state of the ferromagnet at spatial infinity:
$$\nb m(\nb x)\mathop{\longrightarrow}\limits_{|\nb x|\to\infty}
\nb m_0=(0, 0, 1).\eqno (3.4)$$
In the absence of boundaries the medium extends to infinity in all directions
and vortex lines are closed curves. One may then define a degree of
knottedness, or helicity, of tangled vortex lines by analogy with related
work in fluid dynamics [9] and magnetohydrodynamics [10]. The current status
of the topological aspects of the above subjects may be traced from Ref.
[11]. In the present context, such a degree is more appropriately referred
to as the Hopf index [12]. In view of the boundary condition (3.4) the 3D
space is isomorphic to the sphere $S^3$ and a specific configuration
$\nb m=\nb m(\nb x)$ establishes a map from $S^3$ to $S^2$, where $S^2$
is the 2D sphere defined from the constraint $\nb m^2=1$. Such a map is
characterized by the integer-valued Hopf index defined as follows. Let
$\nb m(\nb x)=\nb m_1$ and $\nb m(\nb x)=\nb m_2$ be any two vortex lines
where $\nb m_1$ and $\nb m_2$ are constant unit vectors. The linking number
of these two curves is independent of the specific choice of the pair of
vortex lines and is called the Hopf index of configuration $\nb m(\nb x)$.

In order to make a first contact with the dynamics we also quote an
analytical definition of the Hopf index. The solenoidal vorticity is
derived from a vector potential $\nb a$,
$$\hbox{\bf ã}=\bnabla\times\nb a,\eqno (3.5)$$
and the Hopf index is given by
$$N={1\over 4\pi}\int (\nb a\cdot\hbox{\bf ã})dV.\eqno (3.6)$$
Although the vector potential is unique only to within a gauge
transformation, $N$ is gauge invariant and may be expressed entirely in
terms of $\hbox{\bf ã}$ by
$$N={1\over (4\pi)^2}\int\varepsilon_{ijk}\gamma_i(\nb x)
{(\nb x-\nb x')_j\over |\nb x-\nb x'|^3}\gamma_k(\nb x')dVdV'.\eqno (3.7)$$
The remarkable fact is that the above integral is always equal to an
integer, and its explicit values coincide with those obtained through
the linking-number definition given in the preceding paragraph [13, 14].

It would appear that a simpler (local) expression for the vector potential
may be derived from Eq. (3.3) which suggests that
$$\nb a=\Pi\bnabla\Phi ,\eqno (3.8)$$
provided that $\Pi$ and $\Phi$ are differentiable functions of position.
However inserting (3.3) and (3.8) in (3.6) would then lead to a
vanishing Hopf index. Putting it differently, the canonical variables $\Pi$
and $\Phi$ cannot be both differentiable for field configurations with
$N\not =0$, even though the magnetization is always assumed
to be differentiable. Indeed explicit examples worked out in the literature
[4, 15] demonstrate that the magnetization reaches the north as well as the
south pole of the sphere $\nb m^2=1$ along certain (vortex) lines where the
angular variable $\Phi$ becomes multivalued when $N\not =0$. While these
difficulties are largely irrelevant, because of the gauge-invariant
definition (3.7), they already provide an important hint concerning dynamics.
Note that the vector potential (3.8) coincides with the familiar expression
for the momentum density associated with the Hamilton equations (2.36). We
thus conclude that the difficulties discussed in connection with Eq. (3.8)
render ambiguous also the usual linear momentum.

We defer for the moment further discussion of dynamics and return to the
question of topological classification in the presence of boundaries.
Specifically we consider the film geometry of Fig. 1 where vortex lines need
not be closed but may terminate at the boundaries of the film. Hence a
definition of a Hopf index is no longer meaningful. Instead we consider
the flux of vorticity
$$Q={1\over 4\pi}\int\limits_{S}\hbox{\bf ã}\cdot d\nb S\eqno (3.9)$$
through any open surface $S$ that is contained within the film but extends
to infinity on all sides. The flux is independent of the specific
choice of a surface with the above properties, thanks to
$\bnabla\cdot\hbox{\bf ã}=0$ and an elementary application of the
divergence theorem. In particular, $S$ may be a plane perpendicular to
the third axis,
$$Q={1\over 4\pi}\int\gamma_3dx_1dx_2,\qquad
-{d\over 2}<x_3<{d\over 2},\eqno (3.10)$$
where the double integral is independent of $x_3$. In fact, this integral
coincides with the Pontryagin index or winding number [13] of the
magnetization and is also integer-valued $(Q=0, \pm 1, \pm 2, \ldots)$.
Again, when $Q\not =0$, the canonical variables $\Pi$ and $\Phi$ cannot be
defined everywhere and the corresponding linear momentum is ambiguous.

To be sure, the topological classification described above does not assume
that the configuration $\nb m=\nb m(\nb x)$ solves the Landau-Lifshitz
equation but merely that it obeys some general physical restrictions such as
differentiability and Eq. (3.4). Of course, this classification would become
especially relevant if stationary solutions were found with a nontrivial
topology. In this respect, we note that arguments of varying completeness
have been presented in the literature for the existence of magnetic vortex
rings with a nonvanishing Hopf index [4, 15] but a definite theoretical
treatment and actual observation are still lacking. Nonetheless magnetic
bubbles with a wide range of winding numbers have been observed in
ferromagnetic films [2, 3]. Finally, if the magnetization itself is allowed
to be nondifferentiable at isolated singular points, one is naturally led to
a class of topological {\sl defects} that are also characterized by a winding
number of the form (3.9) except that the surface $S$ is now closed around
a singular point. Such defects have been observed in the bulk of the
ferromagnetic continuum and are called Bloch points [2, 3].

We now organize the various hints concerning the connection between topology
and dynamics by considering the time evolution of the vorticity (3.1). An
elementary calculation based on the Landau-Lifshitz equation at vanishing
dissipation, Eq. (2.15), leads to
$$\dot{\gamma}_i=-\varepsilon_{ijk}\partial_j(\nb f\cdot\partial_k\nb m)=
\varepsilon_{ijk}\partial_j\tau_k,\eqno (3.11)$$
where
$$\tau_k\equiv -(\nb f\cdot\partial_k\nb m)=
\left({\delta W\over\delta\nb m}\cdot\partial_k\nb m\right)\eqno (3.12)$$
is the ``generalized force density'' that appeared first in the work of
Thiele [1]. We take this calculation one step farther using the formal
argument
$$\int\tau_kdV=\int\left({\delta W\over\delta\nb m}\cdot\partial_k\nb m\right)
dV=\partial_kW=0\eqno (3.13)$$

\noindent
to conclude that $\tau_k$ may be written as a total divergence,
$$\tau_k=\partial_{\ell}\sigma_{k\ell},\eqno (3.14)$$
where $\sigma_{k\ell}$ will be called the stress tensor. Equation (3.11)
then reads
$$\dot{\gamma}_i=\varepsilon_{ijk}\partial_j\partial_{\ell}
\sigma_{k\ell}\eqno (3.15)$$
and proves to be fundamental for our purposes [4].

To complete this line of reasoning we must also supply an explicit
expression for the stress tensor. As a first step we insert in Eq. (3.12)
the effective field $\nb f$ of Eq. (2.13):
$$\eqalign{&\tau_k=\tau_k^e+\tau_k^a+\tau_k^b+\tau_k^m,\cr
\noalign{\medskip}
\tau_k^e &=-(\Delta\nb m\cdot\partial_k\nb m),\qquad
\tau_k^a=\kappa (m_1\partial_km_1+m_2\partial_km_2),\cr
\noalign{\medskip}
\tau_k^b &=-h_b\partial_km_3,\qquad \tau_k^m=-(\nb h\cdot\partial_k\nb m).\cr}
\eqno (3.16)$$
We then search for a tensor
$$\sigma_{k\ell}=\sigma_{k\ell}^e+\sigma_{k\ell}^a+
\sigma_{k\ell}^b+\sigma_{k\ell}^m\eqno (3.17)$$
that must lead to Eq. (3.16) by applying the general relation (3.14). The
first three terms are simply
$$\sigma_{k\ell}^e=w_e\delta_{k\ell}
-(\partial_k\nb m\cdot\partial_{\ell}\nb m),\qquad
\sigma_{k\ell}^a=w_a\delta_{k\ell},\qquad
\sigma_{k\ell}^b=w_b\delta_{k\ell},\eqno (3.18)$$
where $w_e$, $w_a$ and $w_b$ are the energy densities defined in Eqs. (2.19)
and (2.23). The construction of the magnetostatic contribution is
slightly more involved but trial and error leads to
$$\sigma_{k\ell}^m=h_kb_{\ell}-{1\over 2}\nb b^2\delta_{k\ell},\eqno (3.19)$$
where the magnetic induction $\nb b=\nb h+\nb m$ is used mostly as a
notational abbreviation. As usual, it is understood that the magnetic field
in Eq. (3.19) is expressed in terms of the magnetization through the
magnetostatic equations (2.14). A repeated application of these equations
establishes the desired relation
$$\partial_{\ell}\sigma_{k\ell}^m=-(\nb h\cdot\partial_k\nb m)=
\tau_k^m.\eqno (3.20)$$

A notable feature of the derived stress tensor $\sigma_{k\ell}$ is that
all but the magnetostatic contributions are symmetric under exchange
of the indices $k$ and $\ell$. The asymmetry of the last term anticipates
the physical fact that the orbital angular momentum and the total
magnetization are not separately conserved in the presence of the
magnetostatic interaction, as we shall see shortly. A further interesting
property is that the preceding construction applies whether or not
boundaries are present, taking into account that the magnetic induction
$\nb b$ is equal to the magnetic field $\nb h$ outside the ferromagnetic
material where the stress tensor reduces to
$$\sigma_{k\ell}=\sigma_{k\ell}^m=h_kh_{\ell}-{1\over 2}\nb h^2
\delta_{k\ell},\eqno (3.21)$$
which is symmetric and satisfies the continuity equation
$$\partial_{\ell}\sigma_{k\ell}=0\eqno (3.22)$$
by virtue of $\bnabla\times\nb h=0=\bnabla\cdot\nb h$. Equation (3.22)
is consistent with a vanishing Thiele force density outside the material.

\bigskip
\bigskip
\bigskip
\noindent
{\ini IV.\ Conservation laws}
\bigskip
The occurrence of ambiguities in the canonical definition of conservation
laws has already received considerable attention. Slonczweski [16] was
apparently the first to recognize that the usual definition of linear
momentum fails for magnetic bubbles with a nonvanishing winding number.
Haldane [17] and Volovik [18] also addressed the question from different
perspectives. But we believe that a simple as well as complete resolution
of this issue was given only in some recent work [4] where the linear and
angular momentum were expressed as moments of the topological vorticity (3.1).
Since the available studies address strictly 2D and 3D models, our aim here
is to establish unambiguous conservation laws in the context of the
quasi-2D geometry appropriate for the description of ferromagnetic films.

We consider the geometry of Fig. 1 where a film of constant thickness $d$
extends to infinity in the $(x_1, x_2)$ plane and the easy axis is
perpendicular to the film. Therefore the relevant symmetries are (i)
translations in the $(x_1, x_2)$ plane, and (ii) azimuthal rotations around
the third axis. We shall show that the corresponding conservation laws
are the moments
$$I_{\mu}=\int x_{\mu}\gamma_3dV,\qquad
\mu =1\quad\hbox{or}\quad 2,\eqno (4.1)$$
which are related to the linear momentum, and the third component of the
total angular momentum
$$J=\ell +\mu;\qquad \ell ={1\over 2}\int\rho^2\gamma_3dV,\qquad
\mu =\int (m_3-1)dV,\eqno (4.2)$$
where $\rho^2=x_1^2+x_2^2$ and hence the ``orbital'' angular momentum $\ell$
is also expressed as a moment of the vorticity; $\mu$ is the total magnetic
moment along the easy axis, except that we have subtracted the trivial
contribution from the ground state so that $\mu$ is finite and negative. It
should be noted that all volume integrals in Eqs. (4.1) and (4.2) extend
over region I of Fig. 1.

Although the conservation laws quoted above have a similar appearance with
those derived for strictly 2D models [4], a proof of their validity is not
obvious because of potential boundary effects. We consider first the time
evolution of the moments (4.1),
$$\dot{I}_{\mu}=\int x_{\mu}\dot{\gamma}_3dV=\varepsilon_{3jk}
\int x_{\mu}\partial_j\partial_{\ell}\sigma_{k\ell}dV,\eqno (4.3)$$
where we have used the fundamental relation (3.15) applied for $i=3$.
Notation is organized by asserting that Greek indices $\mu , \nu , \ldots$
assume only the two distinct values 1 and 2, corresponding to the two
spatial coordinates $x_1$ and $x_2$, while Latin indices $i, j, \ldots$
assume all three values, as usual. We further introduce the 2D
antisymmetric tensor $\varepsilon_{\mu\nu}$, whose elements are
$\varepsilon_{11}=0=\varepsilon_{22}$ and
 $\varepsilon_{12}=1=-\varepsilon_{21}$, and invoke the summation convention
for repeated indices without exception. Then
$$\dot{I}_{\mu}=\varepsilon_{\nu\lambda}\int x_{\mu}\partial_{\nu}
\partial_{\ell}\sigma_{\lambda\ell}dV=\varepsilon_{\nu\lambda}
\int [\partial_{\nu}(x_{\mu}\partial_{\ell}\sigma_{\lambda\ell})-
\delta_{\mu\nu}\partial_{\ell}\sigma_{\lambda\ell}]dV,\eqno (4.4)$$
where both terms in the integrand are in the form of a total divergence.
Since the film extends to infinity in the $x_1$ and $x_2$ directions, the
first integral vanishes for either $\nu =1$ or 2 provided that the
magnetization exhibits a reasonable behavior at large $x_1$ or $x_2$.
By reasonable we mean that the magnetization approaches its ground-state
value sufficiently fast so that the energy of the configuration is finite.
Then we may write
$$\dot{I}_{\mu}=-\varepsilon_{\mu\lambda}\int
\partial_{\ell}\sigma_{\lambda\ell}dV,\eqno (4.5)$$
where the integrand is also a total divergence but the integral need not
vanish because the Latin index $\ell$ is summed over all three values,
$\ell =1, 2$ and 3, and may lead to a nonvanishing contribution from the
film boundaries, namely
$$\dot{I}_{\mu}=-\varepsilon_{\mu\lambda}\left[\,\,\int\limits_{x_3=d/2}
\!\!\!\sigma_{\lambda 3}dx_1dx_2-\!\!\!\!\!\int\limits_{x_3=-d/2}
\!\!\!\!\!\sigma_{\lambda 3}
dx_1dx_2\right] ,\eqno (4.6)$$
where the tensor elements $\sigma_{\lambda 3}$
are evaluated right inside the boundaries and are certainly not equal to zero.

However we may now return to the explicit form of the stress tensor given
in Section III and apply it for $\sigma_{\lambda 3}$ with $\lambda\not =3$
to obtain
$$\sigma_{\lambda 3}=-(\partial_{\lambda}\nb m\cdot\partial_3\nb m)
+h_{\lambda}b_3,\eqno (4.7)$$
which must be evaluated at the boundaries of the film where
$\partial_3\nb m=0$ on account of the unpinned boundary condition (2.22).
Hence
$$\sigma_{\lambda 3}(x_1, x_2, x_3=\pm d/2)=h_{\lambda}b_3,\eqno (4.8)$$
where we further note that the combination of fields $h_{\lambda}b_3,$ with
$\lambda =1$ or 2, is continuous across the boundaries thanks to the familiar
boundary conditions of magnetostatics. The double integrals in Eq. (4.6)
may thus be evaluated right outside the boundaries where the stress tensor
satisfies the continuity equation (3.22). An application of the
divergence theorem in region II yields
$$0=\int\limits_{\hbox{\se II}}\partial_{\ell}\sigma_{\lambda\ell}dV=
-\!\!\!\!\!\int\limits_{x_3=d/2}\!\!\!\!\!
\sigma_{\lambda 3} dx_1dx_2,\eqno (4.9)$$

\noindent
and a similar relation for region III. The net conclusion is that both
integrals in Eq. (4.6) vanish and
$$\dot{I}_{\mu}=0,\eqno (4.10)$$
which is the desired result. We shall defer discussion of the interesting
physical consequences of Eq. (4.10) until a corresponding result is
obtained for the angular momentum.

The time evolution of the orbital angular momentum is governed again by the
fundamental relation (3.15). The analog of Eq. (4.4) now reads
$$\dot{\ell}={1\over 2}\varepsilon_{\nu\lambda}\int\rho^2\partial_{\nu}
\partial_{\ell}\sigma_{\lambda\ell}dV={1\over 2}\varepsilon_{\nu\lambda}
\int [\partial_{\nu}(\rho^2\partial_{\ell}\sigma_{\lambda\ell})-2x_{\nu}
\partial_{\ell}\sigma_{\lambda\ell}]dV,\eqno (4.11)$$
where the first integral in the last step of Eq. (4.11) vanishes for both
$\nu =1$ and 2:
$$\dot{\ell}=-\varepsilon_{\nu\lambda}\int x_{\nu}\partial_{\ell}
\sigma_{\lambda\ell}dV= \varepsilon_{\nu\lambda}\int [\sigma_{\lambda\nu}-
\partial_{\ell}(x_{\nu}\sigma_{\lambda\ell})]dV.\eqno (4.12)$$
Recalling that the volume integration extends over region I we write
$$\int\limits_{\hbox{\se I}}\partial_{\ell}(x_{\nu}\sigma_{\lambda\ell})dV=
\!\!\!\!\!\int\limits_{x_3=d/2}\!\!\!\!\!
x_{\nu}\sigma_{\lambda 3}dx_1dx_2-\!\!\!\!\!\int\limits_{x_3=-d/2}\!\!\!\!\!
x_{\nu}\sigma_{\lambda 3}dx_1dx_2,\eqno (4.13)$$
where the double integrals may be calculated either above or below the
film surfaces because the tensor elements $\sigma_{\lambda 3}$ given
by Eq. (4.8) are continuous across the boundaries. An argument similar
to that used in Eq. (4.9) then leads to
$$\int\limits_{x_3=d/2}\!\!\!\!\! x_{\nu}\sigma_{\lambda 3}dx_1dx_2=-
\int\limits_{\hbox{\se II}}\sigma_{\lambda\nu}dV,\eqno (4.14)$$
and
$$\int\limits_{x_3=-d/2}\!\!\!\!\! x_{\nu}\sigma_{\lambda 3}dx_1dx_2=
\int\limits_{\hbox{\se III}}\sigma_{\lambda\nu} dV.\eqno (4.15)$$
Therefore Eq. (4.13) may be rewritten as
$$\int\limits_{\hbox{\se I}}
\partial_{\ell}(x_{\nu}\sigma_{\lambda\ell})dV=-\int\limits_{\hbox{\se II}}
\sigma_{\lambda\nu}dV-\int\limits_{\hbox{\se III}}
\sigma_{\lambda\nu}dV,\eqno (4.16)$$
where the right-hand side is symmetric under exchange of the indices
$\nu$ and $\lambda$ because the stress tensor is symmetric outside the film.
Hence inserting Eq. (4.16) in Eq. (4.12) yields a vanishing
contribution, because of the contraction with the antisymmetric tensor
$\varepsilon_{\nu\lambda}$, and
$$\dot{\ell}=\int\limits_{\hbox{\se I}}
\varepsilon_{\nu\lambda}\sigma_{\lambda\nu}dV.\eqno (4.17)$$

To summarize, if the magnetostatic interaction were absent, the stress tensor
would be symmetric in all regions and Eq. (4.17) would lead to a conserved
orbital angular momentum $(\dot{\ell}=0)$. In general, using the complete
stress tensor given in Eqs. (3.18) and (3.19),
$$\varepsilon_{\nu\lambda}\sigma_{\lambda\nu}=\varepsilon_{\nu\lambda}
h_{\lambda}b_{\nu}=\varepsilon_{\nu\lambda}h_{\lambda}m_{\nu},\eqno (4.18)$$
and
$$\dot{\ell}=\int (m_1h_2-m_2h_1)dV,\eqno (4.19)$$
so that the orbital angular momentum is not by itself conserved.

Nevertheless a conservation law is obtained by including the total
magnetic moment $\mu$ of Eq. (4.2) whose time derivative is computed by
applying directly the Landau-Lifshitz equation (2.15) to write
$$\dot{\mu}=\int\dot{m}_3dV=-\int (\nb m\times\nb f)_3 dV.\eqno (4.20)$$
Now taking into account the explicit expression for the effective field
$\nb f$ of Eq. (2.13) we find that the contributions from the
anisotropy and bias fields drop out from Eq. (4.20) and
$$\dot{\mu}=-\int (\nb m\times\Delta\nb m+\nb m\times\nb h)_3dV.\eqno (4.21)$$
To compute the exchange contribution we note that
$$\int (\nb m\times\Delta\nb m)dV=\int\partial_i
(\nb m\times\partial_i\nb m)dV=
\oint (\nb m\times\partial_i\nb m)dS_i,\eqno (4.22)$$
where the surface integral vanishes because of the unpinned boundary
condition (2.22). Therefore
$$\dot{\mu}=-\int (\nb m\times\nb h)_3dV=-
\int (m_1h_2-m_2h_1)dV.\eqno (4.23)$$
Comparing this result with Eq. (4.19) establishes that
$$\dot{J}=0,\eqno (4.24)$$
or that the total angular momentum $J=\ell +\mu$ is conserved.

Having thus demonstrated the validity of the conservation laws (4.1) and
(4.2) we now turn to the discussion of their physical content. We first
note that these conservation laws are free of all ambiguities even for
configurations with a nontrivial topological structure. Suffice it to say
that the potential nondifferentiability of the canonical variables
$\Pi$ and $\Phi$ does not affect Eqs. (4.1) and (4.2) because they are
both expressed in terms of the vorticity which can be calculated directly
from the magnetization through Eq. (3.1). A detailed discussion of
this issue may be found in our earlier work within a strictly 2D
context [4] and applies here with minor modifications. Hence we will simply
list the important points adapted to the present quasi-2D situation.

The conserved moments (4.1) are related to the linear momentum
$\nb p=(p_1, p_2)$ by
$$p_{\mu}=\varepsilon_{\mu\nu}I_{\nu},\qquad \{p_{\mu}, \nb m\} =-
\partial_{\mu}\nb m,\eqno (4.25)$$
where the Poisson-bracket relation establishes that $\nb p$ is indeed the
generator of translations in the $(x_1, x_2)$ plane. However $\nb p$
cannot be interpreted as ordinary momentum for two related reasons.
First, the Poisson bracket of its two components,
$$\{ p_1, p_2\} =-4\pi dQ,\eqno (4.26)$$
does not vanish except for a vanishing winding number. Second, under
translations in the plane, $x_1\to x_1+c_1$ and $x_2\to x_2+c_2$,
the moments transform according to
$$I_{\mu}\to I_{\mu}+4\pi dQc_{\mu},\eqno (4.27)$$
which is a consequence of definition (4.1) and Eq. (3.10). The nontrivial
transformation of the linear momentum (4.25) implied by Eq. (4.27) is
surely an unusual property because one would expect the momentum to remain
unchanged under a rigid translation. Nevertheless the above properties suggest
a formal analogy with the familiar electron motion in a uniform magnetic
field, the role of the latter being played here by the winding number.

Therefore, when $Q\not =0$, a more useful interpretation of the conserved
moments is obtained through the guiding center coordinates
$$R_{\mu}={\displaystyle{\int x_{\mu}\gamma_3dV}\over
\displaystyle{\int\gamma_3 dV}}={I_{\mu}\over 4\pi dQ},\qquad
\mu =1\quad\hbox{or}\quad 2,\eqno (4.28)$$
which are conserved and transform as $(R_1, R_2)\to (R_1+c_1, R_2+c_2)$
under a rigid translation in the plane $(x_1, x_2)\to (x_1+c_1, x_2+c_2)$.
The latter property suggests that the 2D vector $\nb R=(R_1, R_2)$ may
be interpreted as the mean position of a magnetic bubble with $Q\not =0$
in a ferromagnetic film, and its conservation implies that such a
bubble cannot be found in a free translational motion. In other words,
$Q\not =0$ bubbles are always spontaneously pinned or frozen within
the ferromagnetic medium provided that external perturbations are absent;
in analogy with electrons undergoing a 2D cyclotron motion in a uniform
magnetic film, in the absence of electric fields.

The physical meaning of the orbital angular momentum $\ell$ defined in Eq.
(4.2) is also unusual, for it actually provides a measure of the size
of a configuration with $Q\not =0$. More precisely, one may define a
mean squared radius from
$$r^2={\displaystyle{\int [(x_1-R_1)^2+(x_2-R_2)^2]\gamma_3dV}\over
\displaystyle{\int\gamma_3dV}}={\ell\over 2\pi dQ}-\nb R^2,\eqno (4.29)$$
which is directly proportional to $\ell$ when the latter is defined
with respect to the guiding center $(\nb R=0)$. Note that we use the
abbreviated 2D notation $\nb R=(R_1, R_2)$ and
$\nb R^2=R_1^2+R_2^2$. The radius $r$ of Eq. (4.29) plays an important
role in our theoretical development but does not, in general, coincide
with the naive radius at which the third component of the magnetization
vanishes $(m_3=0)$. One should also note that $r$ would be a conserved
quantity in the absence of the magnetostatic interaction because the orbital
angular momentum would then be by itself conserved.

In order to pursue further a meaningful discussion of dynamics, one must first
ascertain the existence of interesting static solutions of the
Landau-Lifshitz equation such as magnetic bubbles with a nonvanishing
winding number; an issue addressed in the following two sections. We shall
return to a more detailed study of the implications of the derived
conservation laws for dynamics in Section VII.

\bigskip
\bigskip
\bigskip
\noindent
{\ini V.\ Virial theorems}
\bigskip
A simple scaling argument due to Derrick [7] leads to a virial relation that
must be satisfied by any finite-energy static solution of a nonlinear
field theory. Since Derrick's relation is mainly used in the literature
to establish the nonexistence of nontrivial static solutions, it is of some
interest to demonstrate how the present theory evades its potential
consequences and leads to the observed wealth of magnetic bubbles with
practically any winding number [2]. However a generalization of the original
scaling argument to the present case is not completely straightforward,
because of the film boundaries, and is given below.

Static solutions are stationary points of the energy functional
$W=W(\nb m)$ provided that the constraint $\nb m^2=1$ is taken into account.
For instance, one may use the spherical variables (2.35) to write
$${\delta W\over\delta\Pi}=0={\delta W\over\delta\Phi},\eqno (5.1)$$
which are the static versions of the Hamilton equations (2.36). In this
section, we shall neither write out nor solve the above equations explicitly
but merely use them to derive some general relations.

For the moment, let us ignore the film boundaries and assume that the
medium extends to infinity in all directions. We may then apply Derrick's
scaling argument in a straightforward fashion. Suppose that
$\Pi =\Pi (\nb x)$ and $\Phi =\Phi (\nb x)$ is a solution of Eqs. (5.1) with
(finite) energy $W=W_e+W_a+W_b+W_m$. The energy of the configuration
$\Pi (\zeta\nb x)$ and $\Phi (\zeta\nb x)$, where $\zeta$ is some constant,
is then given by
$$W(\zeta)={1\over\zeta}W_e+{1\over\zeta^3}(W_a+W_b+W_m).\eqno (5.2)$$
By our hypothesis $\zeta =1$ is a stationary point of $W(\zeta)$ and
thus $W'(\zeta =1)=0$ or
$$W_e+3(W_a+W_b+W_m)=0,\eqno (5.3)$$
which is a virial relation that must be satisfied by any static solution
with finite energy. Since all pieces of the energy are positive definite,
one must conclude from Eq. (5.3) that nontrivial static solutions with
finite energy do not exist in a 3D ferromagnetic continuum without boundaries.

The preceding derivation of virial relation (5.2) is clearly inapplicable
in the presence of boundaries. We thus seek to obtain the analog of this
relation for the film geometry of Fig. 1 by a method that was already
employed in the simpler context of Ref. [4] and leads to a series of virial
theorems, Derrick's relation being the simplest example. An alternative form
of the Thiele force density is given by
$$\tau_k={\delta W\over\delta\Pi}\partial_k\Pi +
{\delta W\over\delta\Phi}\partial_k\Phi =\partial_{\ell}
\sigma_{k\ell}\eqno (5.4)$$
and vanishes for static solutions satisfying Eqs. (5.1). Therefore the
stress tensor satisfies the continuity equation
$$\partial_{\ell}\sigma_{k\ell}=0\eqno (5.5)$$
within the ferromagnetic medium. Recalling that the stress tensor satisfies
the continuity equation outside the film even for time-dependent solutions,
see Eq. (3.22), we conclude that static solutions satisfy Eq. (5.5) everywhere.

A series of virial relations may now be derived by taking suitable moments
of Eq. (5.5) and by a systematic application of the divergence theorem.
The simplest possibility is
$$\int\limits_Vx_j\partial_{\ell}\sigma_{k\ell}dV=0,\eqno (5.6)$$
where the integration extends over some volume $V$ that is left
unspecified for the moment. The divergence theorem then yields
$$\int\limits_V\sigma_{ij}dV=\oint\limits_Sx_j
\sigma_{i\ell}dS_{\ell},\eqno (5.7)$$
where we have effected a trivial rearrangement of indices and $S$
is the surface surrounding the volume $V$. It is understood that the region
of integration is such that the surface $S$ does not cross the film
boundaries because of potential discontinuities that may render the
divergence theorem invalid.

Thus we proceed with an application of Eq. (5.7) in several steps.
First we consider the subset of relations obtained by restricting the indices
$i$ and $j$ to the values 1 or 2. Using our standard convention we write
$$\int\limits_V\sigma_{\mu\nu}dV=\oint\limits_Sx_{\nu}\sigma_{\mu\ell}
dS_{\ell},\eqno (5.8)$$
where $\mu , \nu =1$ or 2, and subsequently apply this relation to each region
I, II or III separately:
$$\eqalign{\int\limits_{\hbox{\se I}}
\sigma_{\mu\nu}dV &=S_{\mu\nu}^+-S_{\mu\nu}^-,\cr
\noalign{\medskip}
\int\limits_{\hbox{\se II}}\sigma_{\mu\nu}dV &=-S_{\mu\nu}^+,\cr
\noalign{\medskip}
\int\limits_{\hbox{\se III}}\sigma_{\mu\nu}dV &=S_{\mu\nu}^-,\cr}\eqno (5.9)$$
where
$$S_{\mu\nu}^{\pm}\equiv\!\!\!\!\!\int\limits_{x_3=\pm d/2}\!\!\!\!\!
x_{\nu}\sigma_{\mu 3}dx_1dx_2=\!\!\!\!\!
\int\limits_{x_3=\pm d/2}\!\!\!\!\! x_{\nu}h_{\mu}b_3dx_1dx_2.\eqno (5.10)$$
Here we have recalled the boundary values of the tensor elements
$\sigma_{\mu 3}$ from Eq. (4.8) which are continuous across each boundary
for $\mu =1$ or 2. In fact, the last two equations in (5.9) coincide with
Eqs. (4.14) and (4.15) obtained in our earlier discussion of
conservation laws because the stress tensor satisfies the continuity
equation outside the film even for time-dependent fields. However the
first equation in (5.9) applies only to static solutions. An immediate
consequence of all three equations is the set of relations
$$\int\limits_{\hbox{\se all volume}}\!\!\!\!\!\sigma_{\mu\nu}dV=0;\qquad
\mu, \nu=1\quad\hbox{or}\quad2,\eqno (5.11)$$
where explicit surface contributions are no longer present.
A special case that emphasizes the role of the magnetostatic interaction is
obtained by contracting both sides of Eq. (5.11) with the 2D antisymmetric
tensor,
$$\int\varepsilon_{\nu\mu}\sigma_{\mu\nu}dV=\int
(m_1h_2-m_2h_1)dV=0,\eqno (5.12)$$
a relation that is consistent with Eqs. (4.19) and (4.23) since both the
orbital angular momentum $\ell$ and the total magnetic moment $\mu$ are
time independent in a static solution.

The absence of explicit surface terms in Eq. (5.11) is not surprising
because scaling arguments of the Derrick variety continue to apply in the
$x_1$ and $x_2$ directions. Specifically Eq. (5.11) may be arrived at also
by performing the linear transformation
$x_1\to\zeta_{11}x_1+\zeta_{12}x_2$ and $x_2\to\zeta_{21}x_1+\zeta_{22}x_2$
in a static solution and by demanding that the resulting energy
$W=W(\zeta)$ be stationary at $\zeta_{11}=1=\zeta_{22}$ and
$\zeta_{12}=0=\zeta_{21}$. However the situation is different when one or
both indices $i, j$ in Eq. (5.7) are equal to 3.

Actually some useful information on the latter case can be obtained
directly from the continuity equation (5.5) which is written as
$$\partial_{\nu}\sigma_{i\nu}+\partial_3\sigma_{i3}=0\eqno (5.13)$$
and implies that the double integrals $\int\sigma_{i3}dx_1dx_2$,
with $i=1$, 2 or 3, are independent of $x_3$ but may assume different
values in regions I, II, or III. In fact, all integrals vanish
outside the film because they can be calculated at large $|x_3|$
where the tensor elements vanish. For $i=\mu =1$ or 2 the integrals vanish
also inside the film thanks to Eq. (4.9):
$$\int\sigma_{\mu 3}dx_1dx_2=0,\qquad \mu =1\quad\hbox{or}\quad 2,
\eqno (5.14)$$
for any $x_3$. On the other hand,
$$\int\sigma_{33}dx_1dx_2=\cases{0, &$\qquad |x_3|>d/2,$\cr
\noalign{\medskip}
s, &$-d/2<x_3<d/2,$\cr}\eqno (5.15)$$
where $s$ is constant throughout the film but need  not vanish. Collecting
the above information we may also write
$$\int\limits_{\hbox{\se all volume}}\!\!\!\!\!\sigma_{33} dV=sd,\eqno (5.16)$$
which should be contrasted with Eq. (5.11) where the right-hand side vanishes.

We have thus derived a number of virial relations that must be satisfied by
any static solution. We have also gathered sufficient information to make
contact with relation (5.3) obtained for a strictly 3D medium. Indeed Eqs.
(5.11) and (5.16) may be combined to yield
$$\int \hbox{tr}\sigma dV=sd,\eqno (5.17)$$
where the integration extends over all volume,
$\hbox{tr}\sigma =\sigma_{11}+\sigma_{22}+\sigma_{33}$ is the trace of the
stress tensor, $s$ the constant defined from Eq. (5.15), and $d$ the
film thickness. The trace is calculated by using the explicit expression
of the stress tensor from Eqs. (3.18) and (3.19):
$$\hbox{tr}\sigma =w_e+3(w_a+w_b)+(\nb h\cdot\nb b)-{3\over 2}\nb b^2,
\eqno (5.18)$$
where $w_e, w_a$ and $w_b$ are the exchange, anisotropy and bias energy
densities. We may further insert in Eq. (5.18) the magnetic induction
$\nb b=\nb h+\nb m$ to write
$$\hbox{tr}\sigma =w_e+3(w_a+w_b)-{1\over 2}\nb h^2+4
\left( -{1\over 2}\nb h\cdot\nb m\right)-{3\over 2}\nb m^2,\eqno (5.19)$$
where the magnetostatic energy density appears both in the form entering
Eq. (2.25) and that of Eq. (2.32), while in the last term we must set
$\nb m^2=1$ within the film and zero outside. Therefore a more explicit
form of Eq. (5.17) reads
$$W_e+3[W_a+W_b+(W_m-W_m^{(0)})]=sd,\eqno (5.20)$$
where we recognize the various pieces of the energy, as in relation (5.3),
and $W_m^{(0)}$ originates in the last term of Eq. (5.19)
and is equal to the magnetostatic energy of the ground state
configuration $\nb m_0=(0, 0, 1)$.

Virial relation (5.20) differs from (5.3) in two significant ways.
First, a surface term appears in the right-hand side which is entirely
due to the film geometry and is generally different from zero. Second,
the (infinite) magnetostatic energy of the ground state, $W_m^{(0)}$,
is subtracted out. Now implicit in the derivation of (5.3) was the
assumption that the magnetostatic field vanishes at large distances,
in all directions, so that the energy $W_m$ is finite. This assumption
is clearly false in a ferromagnetic film because $\nb h=-\nb m_0$ at large
$x_1$ and $x_2$ and thus both $W_m$ and $W_m^{(0)}$ are infinite.
Nevertheless the difference $W_m-W_m^{(0)}$ appearing in Eq. (5.20) is
expected to be finite for reasonable solutions. Furthermore this
difference is no longer positive definite and is, in fact, negative in the
case of magnetic bubbles. Indeed the magnetostatic field favors
expansion of a domain with magnetization opposite to that of the ground
state, which is balanced by the exchange, anisotropy and bias fields
to produce a stable bubble of definite radius [19]. Therefore virial
relation (5.20), unlike (5.3), does not a priori exclude nontrivial static
solutions in a ferromagnetic film, irrespectively of the sign of the
surface contribution in the right-hand side. An explicit example is
worked out in the following section where both $W_m-W_m^{(0)}$ and
$s$ are negative but Eq. (5.20) is verified.

\bigskip
\bigskip
\bigskip
\noindent
{\ini VI.\ The fundamental magnetic bubble}
\bigskip
The construction of static solutions with a nonvanishing winding number is an
issue of significant practical interest and occupied most of the early studies
of magnetic bubbles [2, 3]. Because of the long-range nature of the
mangetostatic field and the related effects of finite film thickness,
writing out the static equations (5.1) explicitly leads to a rather complex
system that is not particularly illuminating. Hence the question was addressed
through approximate solutions in the limit of a large quality factor $\kappa$
[19], variational methods [20], and numerical simulations in the important
special case of the fundamental $(Q=1)$ bubble [21]. However, in order to
illustrate some basic aspects of our theoretical development, we shall
need some detailed information on the profile of a bubble that is not
easily accessible from the early work. We have thus decided to recalculate the
$Q=1$ bubble by a numerical method with a simple physical origin.

Suppose that some initial configuration with a given winding number $Q$ evolves
according to the Landau-Lifshitz equation (2.37) including dissipation. After
a sufficiently long time interval precession effects are suppressed and the
configuration eventually relaxes to a static solution of the Landau-Lifshitz
equation with the same winding number. Since our aim in this section is only
to obtain static solutions, the process may be accelerated using Eq. (2.37)
with a very large dissipation constant $\lambda$. On introducing the
rescaled time variable $\tau =t/\lambda$, the $\lambda\to\infty$ limit of
Eq. (2.37) reads
$${\partial\nb m\over\partial\tau}+\nb m\times (\nb m\times\nb f)=0,\qquad
\nb m^2=1.\eqno (6.1)$$
In view of Eq. (2.38) the energy decreases when the configuration
evolves according to either Eq. (2.37) or its fully dissipative limit (6.1).
The advantage of the latter is that it suppresses transients and leads
to equilibrium with reasonable speed. Of course, the calculated static
solution is independent of the details of the initial configuration provided
that the winding number is kept fixed. Thus the initial condition may be
chosen more or less at convenience and convergence may be improved by
incorporating any a priori information on the expected static solution.

Although the principle of the method is very simple, an efficient solution
of the initial-value problem posed in the preceding paragraph confronts us
with a nontrivial numerical task. Thus at every step of the time evolution
one must solve the Poisson equation (2.27) in order to determine the
magnetostatic field $\nb h$ and subsequently the effective field $\nb f$ from
Eq. (2.13). Calculation of the latter near the film boundaries should also
take into account the unpinned boundary condition (2.22). A detailed
description of our numerical algorithm will be given elsewhere [22], so the
remainder of this section will be devoted to a discussion of the results
of an explicit calculation of the fundamental magnetic bubble.

A substantial simplification occurs in the case of the fundamental bubble
because of its {\sl strict} axial symmetry; that is, {\sl invariance}
under a simultaneous rotation in the $(x_1, x_2)$ plane and a
corresponding azimuthal rotation of the magnetization. It is then
convenient to use cylindrical coordinates defined from
$$x_1=\rho\cos\phi ,\qquad x_2=\rho\sin\phi ,\qquad x_3=z.\eqno (6.2)$$
A strictly axially symmetric configuration is of the general form
$$m_1+im_2=(m_{\rho}+im_{\phi})e^{i\phi},\qquad
m_3=m_z\eqno (6.3)$$
where the radial $(m_{\rho})$, azimuthal $(m_{\phi})$ and longitudinal
$(m_z)$ components are functions of only $\rho$ and $z$,
$$m_{\rho}=m_{\rho}(\rho , z),\qquad m_{\phi}=m_{\phi}(\rho , z),\qquad
m_z=m_z(\rho , z),\eqno (6.4)$$
while they continue to satisfy the constraint
$$m_{\rho}^2+m_{\phi}^2+m_z^2=1.\eqno (6.5)$$
The dissipative equation (6.1) becomes effectivelly two-dimensional and a
significant simplification of the numerical problem results.

Specifically, when ansatz (6.3) is inserted in Eq. (6.1), the resulting
equation retains the same form except that the three-component vector
$\nb m=(m_1, m_2, m_3)$ is formally replaced by $(m_{\rho}, m_{\phi}, m_z)$
and the effective field $\nb f$ by $(f_{\rho}, f_{\phi}, f_z)$ with
$$\eqalign{f_{\rho} &=\Delta m_{\rho}-{m_{\rho}\over\rho^2}-\kappa m_{\rho}
+h_{\rho},\cr
\noalign{\medskip}
f_{\phi} &=\Delta m_{\phi}-{m_{\phi}\over\rho^2}-\kappa m_{\phi}+h_{\phi},\cr
\noalign{\medskip}
f_z &=\Delta m_z+h_b+h_z,\cr}\eqno (6.6)$$
where the Laplace operator is reduced to
$$\Delta ={\partial^2\over\partial\rho^2}+{1\over\rho}\,
{\partial\over\partial\rho}+{\partial^2\over\partial z^2},\eqno (6.7)$$
$h_b$ is the bias field, and $h_{\rho}$, $h_{\phi}$ and $h_z$ are the
polar components of the magnetostatic field. Actually, the azimuthal
component vanishes because
$$\bnabla\cdot\nb m={\partial m_{\rho}\over\partial\rho}+
{m_{\rho}\over\rho}+{\partial m_z\over\partial z}\eqno (6.8)$$
and hence the magnetostatic potential is a function of only $\rho$ and $z$;
$\psi =\psi (\rho , z)$. Therefore
$$h_{\rho}=-{\partial\psi\over\partial\rho},\qquad
h_{\phi}=0,\qquad h_z=-{\partial\psi\over\partial z},\eqno (6.9)$$

\noindent
and the polar components of the magnetic induction are
$$b_{\rho}=h_{\rho}+m_{\rho},\qquad b_{\phi}=m_{\phi},\qquad
b_z=h_z+m_z.\eqno (6.10)$$

For future reference we also quote some discrete symmetries of the
reduced system of equations. First, given a static solution of the form (6.4),
the configuration
$$m_{\rho}(\rho , z),\qquad -m_{\phi}(\rho , z),\qquad  m_z(\rho , z)
\eqno (6.11)$$
is also a solution. Second, the parity relations
$$\eqalign{m_{\rho}(\rho , z) &=-m_{\rho}(\rho , -z),\qquad
m_{\phi}(\rho , z)=m_{\phi}(\rho , -z),\qquad
m_z(\rho , z)=m_z(\rho , -z),\cr
\noalign{\medskip}
h_{\rho}(\rho , z) &=-h_{\rho}(\rho , -z),\qquad
h_{\phi}=0,\qquad h_z(\rho , z)=h_z(\rho , -z),\cr}\eqno (6.12)$$
are compatible with the evolution equation (6.1). In other words, if Eq.
(6.1) is solved with an initial condition satisfying relations (6.12),
the resulting static solution will satisfy the same relations.

To complete the description of strictly axially symmetric configurations
we return briefly to the conservation laws (4.1) and (4.2).
The relevant third component of the vorticity reduces to
$$\gamma_3={1\over\rho}\, {\partial m_z\over\partial\rho}.\eqno (6.13)$$
Therefore the winding number calculated from Eq. (3.10) is given by
$$Q={1\over 2}\int^{\infty}_0{\partial m_z\over\partial\rho}d\rho
={1\over 2}[m_z(\infty , z)-m_z(0, z)]=1,\eqno (6.14)$$
provided that the magnetization approaches its ground-state value $m_z=1$
at infinity and the value $m_z=-1$ at the origin. We further note the
trivial fact that the moments $I_{\mu}$ vanish and the guiding center
coincides with the origin of the coordinate system. Finally the orbital
angular momentum is computed from Eqs. (4.2) and (6.13),
$$\ell ={1\over 2}\int^{d/2}_{-d/2}dz\int^{\infty}_0
{\partial m_z\over\partial\rho} 2\pi\rho^2d\rho =-
\int^{d/2}_{-d/2}dz\int^{\infty}_0(m_z-1)2\pi\rho d\rho ,\eqno (6.15)$$
where we have performed a partial integration taking into account that
 $m_z=1$ at infinity. We then recognize in the right-hand side of Eq. (6.15)
the total magnetic moment $\mu$ of Eq. (4.2). Hence $\ell =-\mu$ and
$$J=\ell +\mu =0.\eqno (6.16)$$
As expected, the total angular momentum vanishes for a strictly axially
symmetric configuration. A related fact is that the radius $r$ calculated
from Eq. (4.29) with $\nb R=0$ and $\ell =-\mu$ satisfies the relation
$$\mu =-2\pi dr^2,\eqno (6.17)$$

\noindent
which could also be arrived at by considering a crude model of a bubble
where the magnetization points toward the north pole, $\nb m=(0, 0, 1)$,
for $\rho >r$ and toward the south pole, $\nb m=(0, 0, -1)$, for $\rho <r$.

Now, if Eq. (6.1) is solved for an initial condition with strict axial
symmetry and winding number $Q=1$, it will eventually lead to a static
solution with the same symmetry and winding number. A simple choice of the
initial configuration is given by the two-parameter family
$$m_{\rho}=0,\qquad m_{\phi}=\pm\hbox{sech}\, u,\qquad
m_z=\hbox{tanh}\, u,\eqno (6.18)$$
with
$$u=\ln (\rho/\rho_0)+(\rho -\rho_0)/\delta_0,\eqno (6.19)$$
which coincides with the variational ansatz employed by DeBonte [20] treating
the constants $\rho_0$ and $\delta_0$ as variational parameters. The
constant $\rho_0$ is the naive radius of the bubble, i.e., the radius
at which the third component of the magnetization vanishes, while both
 $\rho_0$ and $\delta_0$ provide a measure of the wall width $\delta_w$
in a picture where the bubble is viewed as a curved domain wall:
$${1\over\delta_w}=\left.{du\over d\rho}\right|_{\rho =\rho_0}=
{1\over\rho_0}+{1\over\delta_0}.\eqno (6.20)$$
On this occasion we recall that the width of an ideal (straight) domain
wall in an infinite medium is
$$\Delta_w=\sqrt{{A\over K}}\quad\hbox{or}\quad {1\over\sqrt{\kappa}}
\eqno (6.21)$$
in the original or rationalized units, respectively (see Section II).
Needless to say, for our purposes the constants $\rho_0$ and $\delta_0$
need not be determined variationally because the relaxation algorithm should
lead to the true static solution for any choice of these parameters. However
convergence may be accelerated when configuration (6.18) is as close as
possible to the true bubble.

The description of the initial ansatz is completed noting that the $\pm$
freedom in Eq. (6.18) reflects the discrete symmetry (6.11). The specific
choice of sign in $m_{\phi}$ will be referred to as the polarity of the
bubble, the winding number being the same $(Q=1)$ for either polarity. Finally
configuration (6.18) is independent of $z$ and trivially satisfies the parity
relations (6.12). Therefore the anticipated static solution will satisfy the
same relations, even though it will develop a nontrivial $z$ dependence.

At this point, one must specify the true parameters of the problem, namely
the quality factor $\kappa$, the bias field $h_b$ and the film thickness $d$.
We have aimed at providing an illustration where the bubble radius is
roughly equal to the film thickness and have thus arrived at the specific
values (in rationalized units)
$$\kappa =2,\qquad h_b=0.32,\qquad d=16\Delta_w={16\over\sqrt{\kappa}},
\eqno (6.22)$$
which belong to a parameter regime that is thought to be ideal for the
formation of magnetic bubbles [19]. A possible choice of the variational
parameters in the initial ansatz (6.19) is accordingly given by
$\rho_0=18\Delta_w$ and $\delta_0=1.1\Delta_w$ but it is certainly
not unique. We finally mention that in all of the ensuing graphical
illustrations of the fundamental bubble we invoke a slight departure
from the rationalized physical units introduced in Section II and used
throughout our theoretical development. Thus distances will now be measured
in units of the ideal domain wall width $\Delta_w=1/\sqrt{\kappa}=
1/\sqrt{2}$ in order to emphasize the wall structure of the calculated bubble.
For instance, the film thickness will appear as $d=16$.

The calculated fundamental magnetic bubble is illustrated in several ways.
We mostly describe a $Q=1$ bubble with positive polarity, originating in
the initial ansatz (6.18) with the upper sign in $m_{\phi}$, the results
for negative polarity being inferred from the discrete symmetry (6.11).
Thus in Fig. 2 we display the dependence of the magnetization on the radial
distance $\rho$ at the film center $(z=0)$ and near the upper boundary
$(z=d/2)$; the $\rho$ dependence near the lower boundary $(z=-d/2)$
may be obtained from the parity relations (6.12). The corresponding results
for the magnetic induction are shown in Fig. 3. One should keep in mind
that the magnetostatic field extends beyond the film boundaries, but the
calculated values will not be discussed further in the present paper.

Some important general features of the fundamental bubble are already
apparent in Fig. 2. If we view the bubble as a curved domain wall, the wall
is purely Bloch at the film center $(m_{\rho}=0)$ and nearly N\'{e}el at
the boundaries where the radial component $m_{\rho}$ achieves significant
values while the azimuthal component $m_{\phi}$ is small. A better view of
the situation is obtained by plotting the projection of the magnetization
vector $\nb m$ on the $(x_1, x_2)$ plane in Fig. 4; whereas Fig. 5
illustrates the projection on a plane that contains the easy axis, which
is chosen to be the $(x_1, x_3)$ plane without loss of generality thanks
to the axial symmetry.

The case of a $Q=1$ bubble with negative polarity may be inferred from Eq.
(6.11) applied, for example, to Fig. 4. The reflection $m_{\phi}\to -m_{\phi}$
will reverse the sense of circulation of the magnetization at the film center
but will not significantly affect the picture at the boundaries where
$m_{\phi}$ is small.

Having thus provided an overall view of the fundamental bubble, we now
turn to the description of some important details. The simplest way to
analyze the fine structure of the bubble is by recalling the concept of
a vortex line introduced in Section III. We first restate the condition of
strict axial symmetry in terms of the spherical variables (2.33):
$$\Theta =\theta (\rho , z),\qquad \Phi =\phi +\chi (\rho , z),\eqno (6.23)$$
where the functions $\theta$ and $\chi$ are independent of the angle $\phi$
and are related to the polar components of the magnetization by
$$\cos\theta =m_z,\qquad \chi =\hbox{arctan} (m_{\phi}/m_{\rho}).\eqno (6.24)$$
Therefore a vortex line is equivalently defined as the intersection of the
two surfaces
$$m_z(\rho , z)=m_3,\qquad \phi +\chi (\rho , z)=\phi_0,\eqno (6.25)$$
where $m_3$ and $\phi_0$ are arbitrary constants in the intervals
$[-1, 1]$ and $[0, 2\pi]$, respectively.

The first relation in (6.25) defines a curve in the $(\rho , z)$ plane,
illustrated in Fig. 6 for three typical values of $m_3$, and the surface
obtained by a simple revolution of the curve around the third axis has
the shape of a barrel. Of special interest is the case $m_3=0$ which
may be used to define a (naive) radius of the bubble $\rho_0=\rho_0(z)$
as the radius of the circular intersection of the barrel with the
$(x_1, x_2)$ plane at altitude $z$. We use the same symbol for the naive
radius as for the variational parameter $\rho_0$ in Eq. (6.19) because the
two coincide within the initial ansatz. The current definition of the
radius is especially useful at the film boundaries where $\rho_0$ is the
distance from the center of the bubble at which a sharp change in contrast
takes place (see Fig. 4) that may be detected experimentally. In our
numerical example we found that $\rho_0(z=\pm d/2)=15.9\Delta_w$ which
should be compared with the value at the film center
 $\rho_0(z=0)=16.3\Delta_w$, thus providing a measure of bubble bulging
[19]. We also quote the average value of the naive radius
$$\bar{\rho}_0={1\over d}\int^{d/2}_{-d/2}\rho_0(z)dz=16.15\Delta_w
\eqno (6.26)$$
and compare it with the radius $r$ defined by Eq. (4.29) and related
to the total magnetic moment by Eq. (6.17):
$$r=16.21\Delta_w.\eqno (6.27)$$
The radius $r$ appears naturally within the theoretical development, as will
become evident in the discussion of skew deflection in Section VII, whereas
the naive radius $\rho_0$ is closer to what is actually measured in an
experiment. Therefore the observed proximity of the numerical values
quoted in Eqs. (6.26) and (6.27) is of special significance. Note
incidentally that the calculated bubble radius for the specific parameters
(6.22) is approximately equal to the film thickness.

Next we consider the second relation in (6.25). We will not attempt to draw
the corresponding surfaces, for various values
 of the constant $\phi_0$, but examine
directly their intersections (vortex lines) with a barrel at given $m_3$.
Thus in Fig. 6 we also display the $z$-dependence of $\chi$ along a
vortex line; that is, we plot the function $\chi (\rho (z; m_3), z)\equiv
\chi (z; m_3)$ where $\rho (z; m_3)$ is the root of the algebraic equation
$m_z(\rho , z)=m_3$ which depends on $z$ and the particular value of $m_3$.
At $m_3=0$, the root $\rho (z; m_3=0)$ reduces to the naive radius
$\rho_0 (z)$ discussed in the preceding paragraph. Actually Fig. 6 shows
the $z$-dependence of $\chi$ along a vortex line only for $m_3=0$, but our
numerical simulation furnished values for $\chi$ that are virtually
indistinguishable when $m_3$ varies in the region $|m_3|< 1/2$.
Nevertheless departures from such a universal behavior occur for
$|m_3|>1/2$.

The preceding findings are best summarized by the sketch of a typical
vortex line given in Fig. 7. Since the sum
$\phi +\chi (\rho (z; m_3), z)$ must remain equal to a constant $\phi_0$,
a vortex line originating at a point $A$ of the upper boundary, i.e.,
at given $\phi_0$ and $m_3$, proceeds downward along the surface of the
barrel at the same time twisting by an amount determined by the variation
of $\chi$ illustrated in the lower part of Fig. 6. The vortex line eventually
terminates at the lower boundary, at a point $B$, having suffered a total
twist $\Delta\phi =-\Delta\chi =\chi (z=d/2; m_3)-\chi (z=-d/2; m_3)$
which generally depends on $m_3$. For the specific example $m_3=0$ shown
in Fig. 6 the calculated total twist is $\Delta\phi =158^{\circ}$.
This value remains practically the same in the range
$|m_3|< 1/2$ but deviations do occur for $|m_3|>1/2$. Finally we note
that the vortex lines twist around the surface of a barrel
counter-clockwise for the $Q=1$ bubble with positive polarity considered
in our illustrations; the twist takes place clockwise for the $Q=1$
bubble with negative polarity obtained through the discrete symmetry (6.11).

Returning to the topological classification discussed in Section III, we note
that all vortex lines in the fundamental magnetic bubble terminate
at the film boundaries and do not tangle with each other. Hence a definition
of a Hopf index is not possible, as expected for a film of finite thickness.
On the other hand, the flux of vorticity through the plane $(x_1, x_2)$
is equal to $4\pi$ for all $z$, thus leading to a unit winding number.

We also return to the virial relations derived in Section V and comment
on the manner they are satisfied in our explicit calculation of the
fundamental bubble. Because of the strict axial symmetry the surface
integrals $S_{\mu\nu}^{\pm}$ of Eq. (5.10) reduce to
$$S_{\mu\nu}^{\pm}=S^{\pm}\delta_{\mu\nu}, \eqno (6.28)$$
where $\delta_{\mu\nu}$ is the 2D Kronecker delta and
$$\eqalignno{S^{\pm}={1\over 2}&\,\,\int^{\infty}_0\rho h_{\rho}b_z
2\pi \rho d\rho .&(6.29)\cr
{\scriptstyle\,\,\,\, x_3}&{\scriptstyle\, =\, \pm d/2}\cr}$$
Hence relations (5.9) applied for $\mu\not =\nu$ simplify to
$$\int\sigma_{12}dV=0=\int\sigma_{21}dV,\eqno (6.30)$$
which are valid in each region I, II or III separately. In fact, using
the explicit expression of the stress tensor for an axially symmetric
configuration, Eqs. (6.30) reduce to the single equation
$$\int h_{\rho}m_{\phi}dV=0,\eqno (6.31)$$
which is automatically satisfied thanks to the parity relations (6.12).
Furthermore Eqs. (5.9) applied for $\mu =\nu =1$ or 2 yield for the
various regions
$$\eqalign{\int\limits_{\hbox{\se I}}\sigma_{11}dV &=\int\limits_{\hbox{\se I}}
\sigma_{22}
dV=S^+-S^-,\cr
\noalign{\medskip}
\int\limits_{\hbox{\se II}}\sigma_{11}dV &=
\int\limits_{\hbox{\se II}}\sigma_{22}
dV=-S^+,\cr
\noalign{\medskip}
\int\limits_{\hbox{\se III}}\sigma_{11}dV &=
\int\limits_{\hbox{\se III}}\sigma_{22}
dV=S^-,\cr}\eqno (6.32)$$
as well as
$$\int\sigma_{11}dV=0=\int\sigma_{22}dV,\eqno (6.33)$$
when the integration extends over all volume. The above relations were
verified explicitly in our numerical simulation and were thus used
to test its validity. We further verified the Derrick-like relation
(5.20) in our specific numerical example where $W_e=4134$,
$W_a=6008$, $W_b=11952$, $W_m-W_m^{(0)}=-22573$, and $sd=-9713$;
here energy is measured in the rationalized units introduced in Section II.

\bigskip
\bigskip
\bigskip
\noindent
{\ini VII.\ Skew deflection}
\bigskip
We are thus ready to study the main dynamical question posed in the
Introduction. A static bubble with winding number $Q$ initially located
at, say, the origin of the coordinate system is subjected to an external
magnetic field
$$\nb h_{\hbox{\se ext}}=(0, 0, h_{\hbox{\se ext}}),\qquad
h_{\hbox{\se ext}}=h_{\hbox{\se ext}} (\nb x, t),\eqno (7.1)$$
that points along the easy axis and its strength is some prescribed
function of position and time. Our task is to determine the response
of the bubble to such an external probe. In the absence of dissipation
$(\lambda =0)$ the relevant dynamical equation is Eq. (2.15) extended
according to
$$\nb f\to\nb f+\nb h_{\hbox{\se ext}},\eqno (7.2)$$
in order to include the effect of the applied field (7.1) which is turned
on at $t=0$. One must then solve the resulting equation with an initial
condition provided by the static bubble and calculate the magnetization
$\nb m=\nb m(\nb x, t)$ at all later times.

However a great deal can be learned without actually solving this
initial-value problem thanks to the special nature of the conservation laws
derived in Section IV. Since the position of the guiding center
$\nb R=(R_1, R_2)$ is conserved in the absence of dissipation and
external fields other than a uniform bias field, examining the rate at
which $\nb R$ changes in the presence of the field (7.1) should yield direct
information on the response of the bubble. Hence the vorticity $\hbox{\bf ã}$
now obeys the relation
$$\dot{\gamma}_i=\varepsilon_{ijk}[\partial_j\partial_{\ell}
\sigma_{k\ell}-\partial_j(\nb h_{\hbox{\se ext}}\cdot\partial_k\nb m)],
\eqno (7.3)$$
which is Eq. (3.11) or (3.15) extended according to Eq. (7.2). In particular,
the evolution of the third component of the vorticity is governed by
$$\dot{\gamma}_3=\varepsilon_{\nu\lambda}[\partial_{\nu}\partial_{\ell}
\sigma_{\lambda\ell}-\partial_{\nu}(\nb h_{\hbox{\se ext}}\cdot
\partial_{\lambda}\nb m],\eqno (7.4)$$
where we have returned to the 2D notation for Greek indices, as in Section IV,
except for the Latin index $\ell$ that is summed over all three values.
The evolution of the moments $I_{\mu}$ of Eq. (4.1) is then given by
$$\dot{I}_{\mu}=\int\varepsilon_{\nu\lambda}x_{\mu}
[\partial_{\nu}\partial_{\ell}\sigma_{\lambda\ell}-\partial_{\nu}
(\nb h_{\hbox{\se ext}}\cdot\partial_{\lambda}\nb m]dV,\eqno (7.5)$$
where the contribution of the first term may be shown to vanish by
reasoning completely analogous to that used in the derivation of the
conservation laws in Section IV. Implicit in the above statement is the
assumption that the applied field (7.1) does not affect significantly the
configuration of the bubble at large distances or, equivalently, it does
not affect the ground state of the ferromagnet. We shall return to this
assumption later in this section. Thus the evolution of the moments in
the presence of the applied field is governed by
$$\dot{I}_{\mu}=-\int\varepsilon_{\nu\lambda}x_{\mu}\partial_{\nu}
(h_{\hbox{\se ext}}\partial_{\lambda}m_3)dV=\int
\varepsilon_{\mu\nu}h_{\hbox{\se ext}}\partial_{\nu}m_3dV,\eqno (7.6)$$
where we have taken into account that the field (7.1) points in the
third direction and have also performed an elementary partial integration.
On the assumption that $m_3\to 1$ sufficiently fast at spatial infinity,
one may perform a further partial integration to write
$$\dot{I}_{\mu}=-\int (\varepsilon_{\mu\nu}\partial_{\nu}
h_{\hbox{\se ext}})(m_3-1)dV.\eqno (7.7)$$
Also recall that the winding number is conserved even in the presence of the
applied field provided that the latter does not destroy the ground state of
the ferromagnet. Therefore the drift velocity of the bubble may be inferred
from Eqs. (4.28) and (7.7):
$$V_{\mu}\equiv\dot{R}_{\mu}=-{1\over 4\pi dQ}\int
(\varepsilon_{\mu\nu}\partial_{\nu}h_{\hbox{\se ext}})
(m_3-1)dV.\eqno (7.8)$$
This result for the drift velocity is not completely explicit because the third
component of the magnetization appearing under the integral sign must still
be determined through a detailed solution of the initial-value problem
described in the introductory paragraphs of this section. However Eq. (7.8)
already contains the essential information concerning the experimentally
observed skew deflection of magnetic bubbles, for it suggests that the drift
velocity will acquire a significant component mainly in a direction
perpendicular to the gradient of the applied field.

In order to appreciate the physical content of Eq. (7.8) the applied field
is written as
$$h_{\hbox{\se ext}}=gx_1,\qquad g=g(\nb x, t),\eqno (7.9)$$
where the ``gradient'' $g$ may still be a function of position and time.
The two components of the drift velocity (7.8) are given equivalently by
$$V_1=-{1\over 4\pi dQ}\int
(x_1\partial_2g)(m_3-1)dV,\qquad V_2={1\over 4\pi dQ}\int
(g+x_1\partial_1g)(m_3-1)dV.\eqno (7.10)$$
The field is now restricted to the physically interesting situation where
the gradient is nearly spatially uniform, i.e. $g\approx g(t)$,
over a large region surrounding the bubble and drops to zero outside that
region. Under such conditions all implicit assumptions made in deriving
Eqs. (7.10) are satisfied. In particular, all partial integrations performed
on the assumption that the field does not significantly alter the behavior of
the bubble at large distances are justified. Nevertheless it is clear from
Eqs. (7.10) that an especially transparent result would be obtained in the
ideal limit where the gradient $g$ is spatially uniform everywhere. Then
$$g=g(t);\qquad V_1=0,\qquad V_2={\mu g\over 4\pi dQ},\eqno (7.11)$$
where $\mu$ is the total magnetic moment defined in Eq. (4.2).

A completely uniform gradient would imply an infinite field at
$x_1\to -\infty$ opposing the magnetization in its ground state
$\nb m=(0, 0, 1)$. Therefore, in the presence of some dissipation, the
magnetization would align with the applied field almost immediately after
the field is turned on and assume the value $\nb m=(0, 0, -1)$ far in the
left plane. Such an instance would destroy the original topological
structure of the bubble and obscure the question of skew deflection.
Nevertheless the preceding criticism does not apply in the case of
vanishing dissipation considered so far because the magnetization would
then precess wildly around the easy axis far in the left plane but never
align  with the external field. Hence the uniform limit considered in Eq.
(7.11) is mathematically meaningful in the absence of dissipation and
provides the clearest, albeit idealized, illustration of the main theme
discussed in this paper.

Thus the guiding center of a bubble with $Q\not =0$ moves in a direction
perpendicular to the applied gradient, in analogy with the familiar Hall
motion of an electric charge in a uniform magnetic field (the analog of the
winding number) and a uniform electric field (the analog of the uniform
magnetic-field gradient). Furthermore the drift velocity (7.11) is
expressed in terms of quantities with a simple physical meaning. However
this expression for the drift velocity is not completely explicit because
the total moment is not conserved during the application of the gradient
and thus acquires some time dependence $\mu =\mu (t)$ that can be determined
only through a detailed solution of the initial-value problem. The moment
$\mu$ would be conserved if the magnetostatic field were absent. Indeed
the orbital angular momentum $\ell$ and the total magnetic moment $\mu$ would
then be separately conserved in the absence of the applied field and, while
the latter violates conservation of $\ell$ because it breaks rotational
symmetry, it does not affect $\mu$ because it points in the third direction.
Under such conditions Eq. (7.11) provides an explicit expression for the
drift velocity since the conserved moment $\mu$ could then be calculated from
the initial configuration of the (static) bubble. This situation occurs in
the case of the 2D isotropic Heisenberg model where an analytical result
for the drift velocity was given in Ref. [4] and was later verified by a
numerical simulation in Ref. [5].

Returning to the realistic case where the magnetostatic field is not
negligible, we note that the total moment $\mu$ may still be calculated
from the profile of the static bubble during the initial stages of the
process. If we further restrict our attention to the fundamental
bubble calculated in Section VI, the moment is related to the bubble radius
through Eq. (6.17) and the initial drift velocity may be written as
$$Q=1;\qquad V_1=0,\qquad V_2=-{1\over 2}gr^2,\eqno (7.12)$$
where we may substitute the numerical value for the bubble radius given
in Eq. (6.27) which is approximately equal to the naive radius measured in
an experiment. Now applying Eq. (7.12) for the speed $V=|V_2|$ yields
$${gr^2\over 2V}=1,\eqno (7.13)$$
which is golden rule (1.1) with a deflection angle $\delta =90^{\circ}$,
as is appropriate in the absence of dissipation, and a winding number
$Q=1$.

However one should keep in mind that the above result provides only a
partial verification of the golden rule for two reasons. First, Eq. (7.13)
is in general violated during the late stages of the process because neither
$\mu$ nor $r$ are conserved; in particular, the relation $\mu =-2\pi dr^2$
is strictly valid only for a static bubble. Second, one must examine the
extent to which (1.1) is valid in the presence of dissipation when the
deflection angle is no longer equal to $90^{\circ}$. The last statement
follows from the simple physical fact that dissipation induces a tendency
for alignment of the magnetization with the external field, which drives
the bubble also toward the left half plane where the field points along
the negative third direction.

In order to study the effect of dissipation more precisely we return to
Eq. (2.37) and extend it according to Eq. (7.2) to write
$$\dot{\nb m}+(\nb m\times\nb G)=0,\qquad \nb m^2=1,\eqno (7.14)$$
where the effective field $\nb G$ is given by
$$\eqalign{\nb G &=\lambda_1(\nb f+\nb h_{\hbox{\se ext}})+\lambda_2
[\nb m\times (\nb f+\nb h_{\hbox{\se ext}})],\cr
\noalign{\medskip}
\lambda_1 &={1\over 1+\lambda^2},\qquad
 \lambda_2={\lambda\over 1+\lambda^2}.\cr}\eqno (7.15)$$
Since Eq. (7.14) is formally identical to the dissipationless equation,
with the replacement $\nb f\to\nb G$, the time derivative of the vorticity
may be inferred from Eq. (3.11) with the same replacement:
$$\dot{\gamma}_i=-\varepsilon_{ijk}\partial_j(\nb G\cdot\partial_k\nb m).
\eqno (7.16)$$
Substitution of the field $\nb G$ from Eq. (7.15) leads to
$$\dot{\gamma}_i=\varepsilon_{ijk}\partial_j
\{\lambda_1\partial_{\ell}\sigma_{k\ell}-\lambda_1
(\nb h_{\hbox{\se ext}}\cdot\partial_k\nb m)-\lambda_2
[\nb m\times (\nb f+\nb h_{\hbox{\se ext}})]\cdot\partial_k\nb m\} ,
\eqno (7.17)$$
a result that may be used to study the time evolution of the guiding center
in a manner analogous to our earlier discussion in the absence of dissipation.
The drift velocity is now given by
$$V_{\mu}={\varepsilon_{\mu\nu}\over 4\pi dQ}\int
[\lambda_1\nb h_{\hbox{\se ext}}+\lambda_2
[\nb m\times (\nb f+\nb h_{\hbox{\se ext}})]\cdot\partial_{\nu}
\nb m dV,\eqno (7.18)$$
which reduces to Eq. (7.8) at vanishing dissipation.

The above result is highly implicit in that the magnetization in the
right-hand side must still be determined through a detailed solution
of the initial value problem. However some explicit information can be
extracted from Eq. (7.18) for the early stages of the bubble motion. The
magnetization may then be calculated from the static profile of the bubble
for which $\nb m\times\nb f=0$. Therefore the initial drift velocity is given
by
$$V_{\mu}={\varepsilon_{\mu\nu}\over 4\pi dQ}\int
[\lambda_1\nb h_{\hbox{\se ext}}+\lambda_2
(\nb m\times\nb h_{\hbox{\se ext}})]\cdot\partial_{\nu}
\nb m dV,\eqno (7.19)$$
where it is understood that the magnetization is that of a static bubble
with winding number $Q$. Taking into account that the field points in the
third direction and performing a partial integration in the first term
yields the equivalent relation
$$V_{\mu}=-{\varepsilon_{\mu\nu}\over 4\pi dQ}\int
[\lambda_1(\partial_{\nu}h_{\hbox{\se ext}})(m_3-1)+\lambda_2h_{\hbox{\se ext}}
(m_1\partial_{\nu}m_2-m_2\partial_{\nu}m_1)]dV.\eqno (7.20)$$

At this point, one should recall the assumptions on the gradient
$g=g(\nb x, t)$ discussed following Eq. (7.9) which are especially
important in the presence of dissipation. Specifically the gradient must
vanish outside a large region surrounding the bubble, for otherwise both
the ground state and the topological structure of the bubble would be
significantly altered. Yet the specific choice of the gradient at large
distances will certainly affect the long-time behavior of the bubble but
should not be crucial during the early motion. Therefore it is still
meaningful to approximate the {\sl initial} drift velocity by inserting in
Eq. (7.20) the applied field $h_{\hbox{\se ext}}=gx_1$ with a gradient
$g=g(t)$ that is spatially uniform. If we further restrict Eq. (7.20)
to the fundamental magnetic bubble calculated in Section VI we find that
$$Q=1;\qquad V_1=-\lambda_2{g\nu\over 4\pi d},\qquad
V_2=\lambda_1{g(\mu +\lambda c)\over 4\pi d},\eqno (7.21)$$
where $\lambda$ is the dissipation constant, $d$ is the film thickness,
and the constants $\mu$, $\nu$ and $c$ are given by
$$\eqalign{\mu =\int (m_z &-1)dV,\qquad
\nu ={1\over 2}\int (m^2_{\rho}+m^2_{\phi})dV,\cr
\noalign{\medskip}
c={1\over 2} &\int\rho \left( m_{\rho} {\partial m_{\phi}\over\partial\rho}-
{\partial m_{\rho}\over\partial\rho}m_{\phi}\right) dV.\cr}\eqno (7.22)$$
Here $\mu$ is the total moment, $\nu$ is essentially the anisotropy energy,
and $c$ vanishes on account of the parity relations (6.12). Therefore our
final result for the initial drift velocity in the presence of dissipation is
$$V_1=-\lambda_2{g\nu\over 4\pi d},\qquad
V_2=\lambda_1{g\mu\over 4\pi d}.\eqno (7.23)$$

Since the total moment $\mu$ is always negative and the constant $\nu$
positive, the guiding center moves off in the lower left plane with an
initial deflection angle $\delta$ with respect to the negative $x_1$ axis
given by
$$\tan\delta ={V_2\over V_1}={1\over\eta\lambda}\quad\hbox{or}\quad
\sin\delta ={1\over\sqrt{1+\eta^2\lambda^2}},\eqno (7.24)$$
where the coefficient
$$\eta =-{\nu\over\mu}=0.08\eqno (7.25)$$
is calculated from Eqs. (7.22) using as input the static bubble. The
specific numerical value quoted above corresponds to the specific choice
of parameters made in Section VI. Finally we calculate the speed
$$V=\sqrt{V^2_1+V^2_2}={\sqrt{1+\eta^2\lambda^2}\over 1+\lambda^2}
{g|\mu|\over 4\pi d}\eqno (7.26)$$
and relate it to the bubble radius through Eq. (6.17) and the deflection
angle through Eq. (7.24):
$${gr^2\over 2V}\sin\delta ={1+\lambda^2\over 1+\eta^2\lambda^2}.\eqno (7.27)$$
At vanishing dissipation $(\lambda =0)$ the deflection angle (7.24) becomes
$\delta =90^{\circ}$ and relation (7.27) reduces to (7.13).

Relation (7.27) establishes contact with the semi-empirical golden rule
(1.1) in the important special case of the fundamental magnetic bubble.
For small values of the dissipation constant $\lambda$ encountered in practice
[2] the right-hand side of Eq. (7.27) is well approximated by unity, to
within terms of order $\lambda^2$, and is thus consistent with Eq. (1.1)
applied for $Q=1$. However this is again only a partial verification of the
golden rule because Eq. (7.27) is strictly valid only for the initial drift
velocity. A complete verification would require first to ascertain that
the bubble eventually reaches a steady state, namely a state with constant
velocity and radius. Such a question could be addressed by a direct
numerical solution of the initial-value problem posed in the first
paragraph of this section. This numerical task is in several respects
similar to the solution of the fully dissipative equation (6.1) described in
Section VI, except for a technical difference that might prove crucial in
practice. Because the applied field breaks rotational invariance the bubble
looses its strict axial symmetry during skew deflection and thus leads
to a 3D numerical simulation that is beyond our current capabilities.

Nevertheless the question was addressed and answered within a strictly 2D
Skyrme model [6] which also leads to Eqs. (7.24) and (7.27) for the initial
drift velocity. However it was found through an explicit numerical solution
of the initial-value problem that a sharp transient period exists during
which the deflection angle departs rapidly and significantly from Eq. (7.24).
The transient period is followed by an intermediate regime where the
deflection angle reaches a more or less constant value and the golden rule
is verified in a rough manner. But a true steady state is never achieved
and the finer predictions of the golden rule are not sustained. In particular,
the long-time behavior of the bubble is sensitive to the details of the
gradient at large distances.

The results from the 2D Skyrme model [6] could be used as a guide for future
numerical investigations of the realistic quasi-2D model studied in the
present paper. We thus turn to a summary of our main conclusions given
in the following section.

\bigskip
\bigskip
\bigskip
\noindent
{\ini VIII.\ Concluding remarks}
\bigskip
We believe to have provided a clear illustration of an important link that
exists between the topological complexity of ferromagnetic structures and
their dynamics. The most direct manifestation of such a link is the
construction of unambiguous conservation laws as moments of the
topological vorticity. The special dynamical features of magnetic bubbles
become transparent and are formally related to more familiar situations
such as the Hall effect of electrodynamics or the Magnus effect of fluid
dynamics.

Our work has also revealed that some of the quantitative predictions of the
early studies must be interpreted with caution. In particular, the golden
rule is valid in its gross features but not in its details. Hence there
exists room for further development of the dynamical theory of magnetic
bubbles. Numerical simulations along the lines of those performed within the
strictly 2D Skyrme model [6] could prove feasible and provide important
hints concerning the remaining questions. There has also been some
speculation to the effect that the dynamics might simplify for hard
$(|Q|\gg 1)$ bubbles, in analogy with the adiabatic dynamics of electric
charges in strong magnetic fields [5]. The semi-empirical golden rule might
then prove to be exact in the extreme large-$Q$ limit and could possibly
be corrected through a systematic adiabatic perturbation theory at finite $Q$.

In this paper we have confined our attention to the response of a bubble
to an externally applied magnetic-field gradient. However a field gradient
is intrinsically present also in the problem of two or more interacting
magnetic bubbles. Thus two interacting bubbles with winding numbers of the
same sign are expected to orbit around each other, in analogy with the 2D
motion of two electrons in a uniform magnetic field or two vortices in an
ordinary fluid. Similarly two bubbles with opposite winding numbers
(e.g., $Q=1$ and $Q=-1$) should move in formation along roughly parallel lines,
also in analogy with an electron-positron pair in a uniform magnetic field
or a vortex-antivortex pair in a fluid. These expectations were confirmed
through numerical simulations in the Skyrme model [6] and should be
possible to establish both theoretically and experimentally in real
ferromagnetic films. Analogous results are expected for interacting
Abrikosov vortices in a superconductor [23].

Finally we return briefly to the possibility of genuinely 3D magnetic
solitons alluded to in Section III. The experimentally observed Bloch points
are 3D topological defects whose dynamics has not yet been studied within
the present framework. Furthermore theoretical arguments for the existence
of magnetic vortex rings with a nonvanishing Hopf index [4, 15] have not
yet been concluded to a definite calculation that would provide the
necessary background for a corresponding experimental search in the bulk
of the ferromagnetic medium.

\bigskip
\bigskip
\noindent
{\bf Acknowledgments}
\bigskip
We are grateful to P.J. Lalousis and P.N. Spathis for valuable assistance
in the numerical simulation described in Section VI. The work was supported
in part by two grants from the EEC(SCI-CT-91-0705 and
CHRX-CT93-0332).

\vfill
\eject

\bigskip
\noindent
{\ini References}
\bigskip
\item{[1] }A.A. Thiele, Phys. Rev. Lett. 30 (1973) 230; J. Appl. Phys. 45
(1974) 377.
\medskip
\item{[2] }A.P. Malozemoff and J.C. Slonczewski, Magnetic Domain Walls in
Bubble Materials (Academic Press, New York, 1979).
\medskip
\item{[3] }T.H. O'Dell, Ferromagnetodynamics, the Dynamics of Magnetic Bubbles,
Domains and Domain Walls (Wiley, New York, 1981).
\medskip
\item{[4] }N. Papanicolaou and T.N. Tomaras, Nucl. Phys. B 360 (1991) 425;
N. Papanicolaou, in Singularities in Fluids, Plasmas and Optics,
p. 151-158, eds., R.E. Caflisch and G.C. Papanicolaou (Kluwer, Amsterdam,
1993).
\medskip
\item{[5] }N. Papanicolaou, Physica D 74 (1994) 107; Phys. Lett. A 186
(1994) 119.
\medskip
\item{[6] }N. Papanicolaou and W.J. Zakrzewski, Physica D 80 (1995) 225;
and Dynamics of Magnetic Bubbles in a Skyrme Model, preprint (1995).
\medskip
\item{[7] }G.H. Derrick, J. Math. Phys. 5 (1964) 1252.
\medskip
\item{[8] }G.K. Batchelor, An Introduction to Fluid Dynamics (Cambridge
University Press, 1967).
\medskip
\item{[9] }H.K. Moffat, J. Fluid Mech. 35 (1969) 117.
\medskip
\item{[10] }H.K. Moffat, Magnetic Field Generation in Electrically
Conducting Fluids (Cambridge University Press, 1978).
\medskip
\item{[11] }Topological Aspects of the Dynamics of Fluids and Plasmas, NATO
ASI Series, eds., H.K. Moffat, G.M. Zaslavsky, P. Compte and M. Tabor
(Kluwer, Dordrecht, 1992).
\medskip
\item{[12] }E.A. Kuznetsov and A.V. Mikhailov, Phys. Lett. A 77 (1980) 37.
\medskip
\item{[13] }R. Bott and L.W. Tu, Differential Forms in Algebraic Topology
(Springer, New York, 1982).
\medskip
\item{[14] }X.G. Wen and A. Zee, Phys. Rev. Lett. 61 (1988) 1025.
\medskip
\item{[15] }J.E. Dzyaloshinskii and B.A. Ivanov, JETP Lett. 29 (1979) 540.
\medskip
\item{[16] }J.C. Slonczewski, J. Magn. Magn. Mat. 12 (1979) 108.
\medskip
\item{[17] }F.D.M. Haldane, Phys. Rev. Lett. 57 (1986) 1488.
\medskip
\item{[18] }G.E. Volovik, J. Phys. C: Solid State Phys. 20 (1987) L83.
\medskip
\item{[19] }A.A. Thiele, Bell Syst. Tech. J. 48 (1969) 3287;
J. Appl. Phys. 41 (1970) 1139.
\medskip
\item{[20] }W.J. DeBonte AIP Conf. Proc. 5 (1971) 140; J. Appl. Phys. 44
(1973) 1793; IEEE Trans. Magn., MAG-11 (1975) 3.
\medskip
\item{[21] }T.G.W. Blake, E. Della Torre, J. Appl. Phys. 50 (1979) 2192;
E. Della Torre, C. Haged\"{u}s, G. K\'{a}d\'{a}r, AIP Conf. Proc.
29 (1975) 89.
\medskip
\item{[22] }S. Komineas, PhD thesis, University of Crete.
\medskip
\item{[23] }N. Papanicolaou and T.N. Tomaras, Phys. Lett. A 179 (1993) 33.

\vfill
\eject

\noindent
{\ini Figure captions}
\bigskip
\noindent
{\bf Figure 1:} Conventions concerning a ferromagnetic film of thickness $d$.
The film extends to infinity in the $x_1$ and $x_2$ directions.
\bigskip
\noindent
{\bf Figure 2:} The calculated magnetization for the fundamental $(Q=1)$
magnetic bubble with positive polarity and parameters specified by Eq. (6.22).
Results are given at the film center $(z=0)$ and at the upper boundary
$(z=d/2)$, whereas the corresponding results at the lower boundary
$(z=-d/2)$ may be inferred from the parity relations (6.12). Here and in
all subsequent graphical illustrations distance is measured in units of the
ideal wall width $\Delta_w=1/\sqrt{\kappa}$.
\bigskip
\noindent
{\bf Figure 3:} The calculated magnetic induction for the $Q=1$ bubble;
see caption of Fig. 2 for further explanations.
\bigskip
\noindent
{\bf Figure 4:} Illustration of the $Q=1$ bubble with positive polarity
through the projection of the magnetization vector field $\nb m$ on the
$(x_1, x_2)$ plane. The bubble is Bloch-like at the film center
$(x_3=0)$ and N\'{e}el-like near the boundaries $(x_3=\pm d/2)$.
\bigskip
\noindent
{\bf Figure 5:} Another view of the $Q=1$ bubble through the projection
of $\nb m$ on the $(x_1, x_3)$ plane.
\bigskip
\noindent
{\bf Figure 6:} The calculated curves $m_z(\rho , z)=m_3$ for
$m_3=0$, $\pm 1/2$ (upper entry) and the $z$-dependence of $\chi$ along
a vortex line (lower entry) for the $Q=1$ bubble with positive polarity;
see the text for further explanations.
\bigskip
\noindent
{\bf Figure 7:} Sketch of a typical vortex line for a $Q=1$ bubble with
positive polarity. The sense of twist is reversed for a $Q=1$ bubble with
negative polarity.

\end